\documentclass[longauth]{aaEC}
\usepackage[utf8]{inputenc}
\usepackage{graphicx}
\usepackage{natbib}
\usepackage[table]{xcolor}
\usepackage{multirow}
\usepackage{ragged2e}
\usepackage{totcount}
\newtotcounter{citnum}
\def\oldbibitem{} \let\oldbibitem=\bibitem
\def\bibitem{\stepcounter{citnum}\oldbibitem}
\usepackage{txfonts}
\usepackage[pdfencoding=auto,psdextra]{hyperref}
\hypersetup{
    colorlinks=true,
    linkcolor=blue,
    filecolor=magenta, 
    urlcolor=blue,
    citecolor=blue
}
\makeatletter
\renewcommand*\aa@pageof{, page \thepage{} of \pageref*{LastPage}}
\makeatother
\usepackage[utf8]{inputenc}
\usepackage[switch, modulo]{lineno}
\AtBeginDocument{\let\linenumbers\nolinenumbers}
\usepackage{euclid}

\begin{document}

\title{\Euclid\/: A blue galaxy population and a brightest cluster galaxy in the making in a $z\sim 1.74$ MaDCoWS2 galaxy cluster candidate\thanks{This paper is published on behalf of the Euclid Consortium}}
\newcommand{\orcid}[1]{}

\author{A.~Trudeau\orcid{0000-0003-3428-1106}\thanks{\email{atrudeau@asiaa.sinica.edu.tw}}\inst{\ref{aff1},\ref{aff2}}
\and A.~H.~Gonzalez\orcid{0000-0002-0933-8601}\inst{\ref{aff2}}
\and S.~A.~Stanford\orcid{0000-0003-0122-0841}\inst{\ref{aff3}}
\and S.~Shamyati\orcid{0009-0007-4472-6136}\inst{\ref{aff4}}
\and S.~Taamoli\orcid{0000-0003-0749-4667}\inst{\ref{aff4}}
\and D.~Stern\orcid{0000-0003-2686-9241}\inst{\ref{aff5}}
\and P.~R.~M.~Eisenhardt\inst{\ref{aff5}}
\and B.~Mobasher\orcid{0000-0001-5846-4404}\inst{\ref{aff4}}
\and K.~Thongkham\orcid{0000-0001-7027-2202}\inst{\ref{aff6},\ref{aff7},\ref{aff2}}
\and B.~Altieri\orcid{0000-0003-3936-0284}\inst{\ref{aff8}}
\and S.~Andreon\orcid{0000-0002-2041-8784}\inst{\ref{aff9}}
\and C.~Baccigalupi\orcid{0000-0002-8211-1630}\inst{\ref{aff10},\ref{aff11},\ref{aff12},\ref{aff13}}
\and M.~Baldi\orcid{0000-0003-4145-1943}\inst{\ref{aff14},\ref{aff15},\ref{aff16}}
\and A.~Balestra\orcid{0000-0002-6967-261X}\inst{\ref{aff17}}
\and S.~Bardelli\orcid{0000-0002-8900-0298}\inst{\ref{aff15}}
\and A.~Biviano\orcid{0000-0002-0857-0732}\inst{\ref{aff11},\ref{aff10}}
\and E.~Branchini\orcid{0000-0002-0808-6908}\inst{\ref{aff18},\ref{aff19},\ref{aff9}}
\and M.~Brescia\orcid{0000-0001-9506-5680}\inst{\ref{aff20},\ref{aff21}}
\and S.~Camera\orcid{0000-0003-3399-3574}\inst{\ref{aff22},\ref{aff23},\ref{aff24}}
\and G.~Ca\~nas-Herrera\orcid{0000-0003-2796-2149}\inst{\ref{aff25},\ref{aff26}}
\and V.~Capobianco\orcid{0000-0002-3309-7692}\inst{\ref{aff24}}
\and C.~Carbone\orcid{0000-0003-0125-3563}\inst{\ref{aff27}}
\and J.~Carretero\orcid{0000-0002-3130-0204}\inst{\ref{aff28},\ref{aff29}}
\and S.~Casas\orcid{0000-0002-4751-5138}\inst{\ref{aff30},\ref{aff31}}
\and M.~Castellano\orcid{0000-0001-9875-8263}\inst{\ref{aff32}}
\and G.~Castignani\orcid{0000-0001-6831-0687}\inst{\ref{aff15}}
\and S.~Cavuoti\orcid{0000-0002-3787-4196}\inst{\ref{aff21},\ref{aff33}}
\and K.~C.~Chambers\orcid{0000-0001-6965-7789}\inst{\ref{aff34}}
\and A.~Cimatti\inst{\ref{aff35}}
\and C.~Colodro-Conde\inst{\ref{aff36}}
\and G.~Congedo\orcid{0000-0003-2508-0046}\inst{\ref{aff37}}
\and C.~J.~Conselice\orcid{0000-0003-1949-7638}\inst{\ref{aff38}}
\and L.~Conversi\orcid{0000-0002-6710-8476}\inst{\ref{aff39},\ref{aff8}}
\and Y.~Copin\orcid{0000-0002-5317-7518}\inst{\ref{aff40}}
\and F.~Courbin\orcid{0000-0003-0758-6510}\inst{\ref{aff41},\ref{aff42},\ref{aff43}}
\and H.~M.~Courtois\orcid{0000-0003-0509-1776}\inst{\ref{aff44}}
\and M.~Cropper\orcid{0000-0003-4571-9468}\inst{\ref{aff45}}
\and A.~Da~Silva\orcid{0000-0002-6385-1609}\inst{\ref{aff46},\ref{aff47}}
\and H.~Degaudenzi\orcid{0000-0002-5887-6799}\inst{\ref{aff48}}
\and G.~De~Lucia\orcid{0000-0002-6220-9104}\inst{\ref{aff11}}
\and H.~Dole\orcid{0000-0002-9767-3839}\inst{\ref{aff49}}
\and M.~Douspis\orcid{0000-0003-4203-3954}\inst{\ref{aff49}}
\and F.~Dubath\orcid{0000-0002-6533-2810}\inst{\ref{aff48}}
\and C.~A.~J.~Duncan\orcid{0009-0003-3573-0791}\inst{\ref{aff37}}
\and X.~Dupac\inst{\ref{aff8}}
\and S.~Dusini\orcid{0000-0002-1128-0664}\inst{\ref{aff50}}
\and S.~Escoffier\orcid{0000-0002-2847-7498}\inst{\ref{aff51}}
\and M.~Fabricius\orcid{0000-0002-7025-6058}\inst{\ref{aff52},\ref{aff53}}
\and M.~Farina\orcid{0000-0002-3089-7846}\inst{\ref{aff54}}
\and F.~Faustini\orcid{0000-0001-6274-5145}\inst{\ref{aff32},\ref{aff55}}
\and S.~Ferriol\inst{\ref{aff40}}
\and F.~Finelli\orcid{0000-0002-6694-3269}\inst{\ref{aff15},\ref{aff56}}
\and M.~Frailis\orcid{0000-0002-7400-2135}\inst{\ref{aff11}}
\and E.~Franceschi\orcid{0000-0002-0585-6591}\inst{\ref{aff15}}
\and M.~Fumana\orcid{0000-0001-6787-5950}\inst{\ref{aff27}}
\and S.~Galeotta\orcid{0000-0002-3748-5115}\inst{\ref{aff11}}
\and K.~George\orcid{0000-0002-1734-8455}\inst{\ref{aff57}}
\and B.~Gillis\orcid{0000-0002-4478-1270}\inst{\ref{aff37}}
\and C.~Giocoli\orcid{0000-0002-9590-7961}\inst{\ref{aff15},\ref{aff16}}
\and J.~Gracia-Carpio\inst{\ref{aff52}}
\and A.~Grazian\orcid{0000-0002-5688-0663}\inst{\ref{aff17}}
\and F.~Grupp\inst{\ref{aff52},\ref{aff53}}
\and S.~V.~H.~Haugan\orcid{0000-0001-9648-7260}\inst{\ref{aff58}}
\and W.~Holmes\inst{\ref{aff5}}
\and F.~Hormuth\inst{\ref{aff59}}
\and A.~Hornstrup\orcid{0000-0002-3363-0936}\inst{\ref{aff60},\ref{aff61}}
\and K.~Jahnke\orcid{0000-0003-3804-2137}\inst{\ref{aff62}}
\and M.~Jhabvala\inst{\ref{aff63}}
\and B.~Joachimi\orcid{0000-0001-7494-1303}\inst{\ref{aff64}}
\and E.~Keih\"anen\orcid{0000-0003-1804-7715}\inst{\ref{aff65}}
\and S.~Kermiche\orcid{0000-0002-0302-5735}\inst{\ref{aff51}}
\and M.~Kilbinger\orcid{0000-0001-9513-7138}\inst{\ref{aff66}}
\and B.~Kubik\orcid{0009-0006-5823-4880}\inst{\ref{aff40}}
\and M.~K\"ummel\orcid{0000-0003-2791-2117}\inst{\ref{aff53}}
\and M.~Kunz\orcid{0000-0002-3052-7394}\inst{\ref{aff67}}
\and H.~Kurki-Suonio\orcid{0000-0002-4618-3063}\inst{\ref{aff68},\ref{aff69}}
\and A.~M.~C.~Le~Brun\orcid{0000-0002-0936-4594}\inst{\ref{aff70}}
\and D.~Le~Mignant\orcid{0000-0002-5339-5515}\inst{\ref{aff71}}
\and S.~Ligori\orcid{0000-0003-4172-4606}\inst{\ref{aff24}}
\and P.~B.~Lilje\orcid{0000-0003-4324-7794}\inst{\ref{aff58}}
\and V.~Lindholm\orcid{0000-0003-2317-5471}\inst{\ref{aff68},\ref{aff69}}
\and I.~Lloro\orcid{0000-0001-5966-1434}\inst{\ref{aff72}}
\and G.~Mainetti\orcid{0000-0003-2384-2377}\inst{\ref{aff73}}
\and D.~Maino\inst{\ref{aff74},\ref{aff27},\ref{aff75}}
\and E.~Maiorano\orcid{0000-0003-2593-4355}\inst{\ref{aff15}}
\and O.~Mansutti\orcid{0000-0001-5758-4658}\inst{\ref{aff11}}
\and O.~Marggraf\orcid{0000-0001-7242-3852}\inst{\ref{aff76}}
\and M.~Martinelli\orcid{0000-0002-6943-7732}\inst{\ref{aff32},\ref{aff77}}
\and N.~Martinet\orcid{0000-0003-2786-7790}\inst{\ref{aff71}}
\and F.~Marulli\orcid{0000-0002-8850-0303}\inst{\ref{aff78},\ref{aff15},\ref{aff16}}
\and R.~J.~Massey\orcid{0000-0002-6085-3780}\inst{\ref{aff79}}
\and S.~Maurogordato\inst{\ref{aff80}}
\and E.~Medinaceli\orcid{0000-0002-4040-7783}\inst{\ref{aff15}}
\and S.~Mei\orcid{0000-0002-2849-559X}\inst{\ref{aff81},\ref{aff82}}
\and Y.~Mellier\thanks{Deceased}\inst{\ref{aff83},\ref{aff84}}
\and M.~Meneghetti\orcid{0000-0003-1225-7084}\inst{\ref{aff15},\ref{aff16}}
\and E.~Merlin\orcid{0000-0001-6870-8900}\inst{\ref{aff32}}
\and G.~Meylan\inst{\ref{aff85}}
\and A.~Mora\orcid{0000-0002-1922-8529}\inst{\ref{aff86}}
\and L.~Moscardini\orcid{0000-0002-3473-6716}\inst{\ref{aff78},\ref{aff15},\ref{aff16}}
\and E.~Munari\orcid{0000-0002-1751-5946}\inst{\ref{aff11},\ref{aff10}}
\and R.~Nakajima\orcid{0009-0009-1213-7040}\inst{\ref{aff76}}
\and C.~Neissner\orcid{0000-0001-8524-4968}\inst{\ref{aff87},\ref{aff29}}
\and S.-M.~Niemi\orcid{0009-0005-0247-0086}\inst{\ref{aff25}}
\and C.~Padilla\orcid{0000-0001-7951-0166}\inst{\ref{aff87}}
\and S.~Paltani\orcid{0000-0002-8108-9179}\inst{\ref{aff48}}
\and F.~Pasian\orcid{0000-0002-4869-3227}\inst{\ref{aff11}}
\and K.~Pedersen\inst{\ref{aff88}}
\and W.~J.~Percival\orcid{0000-0002-0644-5727}\inst{\ref{aff89},\ref{aff90},\ref{aff91}}
\and V.~Pettorino\orcid{0000-0002-4203-9320}\inst{\ref{aff25}}
\and S.~Pires\orcid{0000-0002-0249-2104}\inst{\ref{aff66}}
\and G.~Polenta\orcid{0000-0003-4067-9196}\inst{\ref{aff55}}
\and M.~Poncet\inst{\ref{aff92}}
\and L.~A.~Popa\inst{\ref{aff93}}
\and L.~Pozzetti\orcid{0000-0001-7085-0412}\inst{\ref{aff15}}
\and F.~Raison\orcid{0000-0002-7819-6918}\inst{\ref{aff52}}
\and A.~Renzi\orcid{0000-0001-9856-1970}\inst{\ref{aff94},\ref{aff50}}
\and J.~Rhodes\orcid{0000-0002-4485-8549}\inst{\ref{aff5}}
\and G.~Riccio\inst{\ref{aff21}}
\and E.~Romelli\orcid{0000-0003-3069-9222}\inst{\ref{aff11}}
\and M.~Roncarelli\orcid{0000-0001-9587-7822}\inst{\ref{aff15}}
\and R.~Saglia\orcid{0000-0003-0378-7032}\inst{\ref{aff53},\ref{aff52}}
\and Z.~Sakr\orcid{0000-0002-4823-3757}\inst{\ref{aff95},\ref{aff96},\ref{aff97}}
\and D.~Sapone\orcid{0000-0001-7089-4503}\inst{\ref{aff98}}
\and B.~Sartoris\orcid{0000-0003-1337-5269}\inst{\ref{aff53},\ref{aff11}}
\and P.~Schneider\orcid{0000-0001-8561-2679}\inst{\ref{aff76}}
\and T.~Schrabback\orcid{0000-0002-6987-7834}\inst{\ref{aff99}}
\and A.~Secroun\orcid{0000-0003-0505-3710}\inst{\ref{aff51}}
\and G.~Seidel\orcid{0000-0003-2907-353X}\inst{\ref{aff62}}
\and S.~Serrano\orcid{0000-0002-0211-2861}\inst{\ref{aff100},\ref{aff101},\ref{aff102}}
\and C.~Sirignano\orcid{0000-0002-0995-7146}\inst{\ref{aff94},\ref{aff50}}
\and G.~Sirri\orcid{0000-0003-2626-2853}\inst{\ref{aff16}}
\and L.~Stanco\orcid{0000-0002-9706-5104}\inst{\ref{aff50}}
\and J.~Steinwagner\orcid{0000-0001-7443-1047}\inst{\ref{aff52}}
\and P.~Tallada-Cresp\'{i}\orcid{0000-0002-1336-8328}\inst{\ref{aff28},\ref{aff29}}
\and A.~N.~Taylor\inst{\ref{aff37}}
\and H.~I.~Teplitz\orcid{0000-0002-7064-5424}\inst{\ref{aff103}}
\and I.~Tereno\orcid{0000-0002-4537-6218}\inst{\ref{aff46},\ref{aff104}}
\and N.~Tessore\orcid{0000-0002-9696-7931}\inst{\ref{aff45}}
\and S.~Toft\orcid{0000-0003-3631-7176}\inst{\ref{aff105},\ref{aff106}}
\and R.~Toledo-Moreo\orcid{0000-0002-2997-4859}\inst{\ref{aff107}}
\and F.~Torradeflot\orcid{0000-0003-1160-1517}\inst{\ref{aff29},\ref{aff28}}
\and I.~Tutusaus\orcid{0000-0002-3199-0399}\inst{\ref{aff102},\ref{aff100},\ref{aff96}}
\and L.~Valenziano\orcid{0000-0002-1170-0104}\inst{\ref{aff15},\ref{aff56}}
\and J.~Valiviita\orcid{0000-0001-6225-3693}\inst{\ref{aff68},\ref{aff69}}
\and T.~Vassallo\orcid{0000-0001-6512-6358}\inst{\ref{aff11}}
\and Y.~Wang\orcid{0000-0002-4749-2984}\inst{\ref{aff108}}
\and J.~Weller\orcid{0000-0002-8282-2010}\inst{\ref{aff53},\ref{aff52}}
\and G.~Zamorani\orcid{0000-0002-2318-301X}\inst{\ref{aff15}}
\and F.~M.~Zerbi\inst{\ref{aff9}}
\and E.~Zucca\orcid{0000-0002-5845-8132}\inst{\ref{aff15}}
\and J.~Garc\'ia-Bellido\orcid{0000-0002-9370-8360}\inst{\ref{aff109}}
\and M.~Maturi\orcid{0000-0002-3517-2422}\inst{\ref{aff95},\ref{aff110}}
\and V.~Scottez\orcid{0009-0008-3864-940X}\inst{\ref{aff83},\ref{aff111}}
\and M.~Sereno\orcid{0000-0003-0302-0325}\inst{\ref{aff15},\ref{aff16}}}

\institute{Academia Sinica Institute of Astronomy and Astrophysics (ASIAA), 11F of ASMAB, No.~1, Section 4, Roosevelt Road, Taipei 10617, Taiwan\label{aff1}
\and
Department of Astronomy, University of Florida, Bryant Space Science Center, Gainesville, FL 32611, USA\label{aff2}
\and
Department of Physics and Astronomy, University of California, Davis, CA 95616, USA\label{aff3}
\and
Physics and Astronomy Department, University of California, 900 University Ave., Riverside, CA 92521, USA\label{aff4}
\and
Jet Propulsion Laboratory, California Institute of Technology, 4800 Oak Grove Drive, Pasadena, CA, 91109, USA\label{aff5}
\and
Korea Astronomy and Space Science Institute, 776 Daedeok-daero, Yuseong-gu, Daejeon 34055, Republic of Korea\label{aff6}
\and
National Astronomical Research Institute of Thailand (NARIT), Mae Rim, Chiang Mai 50180, Thailand\label{aff7}
\and
ESAC/ESA, Camino Bajo del Castillo, s/n., Urb. Villafranca del Castillo, 28692 Villanueva de la Ca\~nada, Madrid, Spain\label{aff8}
\and
INAF-Osservatorio Astronomico di Brera, Via Brera 28, 20122 Milano, Italy\label{aff9}
\and
IFPU, Institute for Fundamental Physics of the Universe, via Beirut 2, 34151 Trieste, Italy\label{aff10}
\and
INAF-Osservatorio Astronomico di Trieste, Via G. B. Tiepolo 11, 34143 Trieste, Italy\label{aff11}
\and
INFN, Sezione di Trieste, Via Valerio 2, 34127 Trieste TS, Italy\label{aff12}
\and
SISSA, International School for Advanced Studies, Via Bonomea 265, 34136 Trieste TS, Italy\label{aff13}
\and
Dipartimento di Fisica e Astronomia, Universit\`a di Bologna, Via Gobetti 93/2, 40129 Bologna, Italy\label{aff14}
\and
INAF-Osservatorio di Astrofisica e Scienza dello Spazio di Bologna, Via Piero Gobetti 93/3, 40129 Bologna, Italy\label{aff15}
\and
INFN-Sezione di Bologna, Viale Berti Pichat 6/2, 40127 Bologna, Italy\label{aff16}
\and
INAF-Osservatorio Astronomico di Padova, Via dell'Osservatorio 5, 35122 Padova, Italy\label{aff17}
\and
Dipartimento di Fisica, Universit\`a di Genova, Via Dodecaneso 33, 16146, Genova, Italy\label{aff18}
\and
INFN-Sezione di Genova, Via Dodecaneso 33, 16146, Genova, Italy\label{aff19}
\and
Department of Physics "E. Pancini", University Federico II, Via Cinthia 6, 80126, Napoli, Italy\label{aff20}
\and
INAF-Osservatorio Astronomico di Capodimonte, Via Moiariello 16, 80131 Napoli, Italy\label{aff21}
\and
Dipartimento di Fisica, Universit\`a degli Studi di Torino, Via P. Giuria 1, 10125 Torino, Italy\label{aff22}
\and
INFN-Sezione di Torino, Via P. Giuria 1, 10125 Torino, Italy\label{aff23}
\and
INAF-Osservatorio Astrofisico di Torino, Via Osservatorio 20, 10025 Pino Torinese (TO), Italy\label{aff24}
\and
European Space Agency/ESTEC, Keplerlaan 1, 2201 AZ Noordwijk, The Netherlands\label{aff25}
\and
Leiden Observatory, Leiden University, Einsteinweg 55, 2333 CC Leiden, The Netherlands\label{aff26}
\and
INAF-IASF Milano, Via Alfonso Corti 12, 20133 Milano, Italy\label{aff27}
\and
Centro de Investigaciones Energ\'eticas, Medioambientales y Tecnol\'ogicas (CIEMAT), Avenida Complutense 40, 28040 Madrid, Spain\label{aff28}
\and
Port d'Informaci\'{o} Cient\'{i}fica, Campus UAB, C. Albareda s/n, 08193 Bellaterra (Barcelona), Spain\label{aff29}
\and
Institute for Theoretical Particle Physics and Cosmology (TTK), RWTH Aachen University, 52056 Aachen, Germany\label{aff30}
\and
Deutsches Zentrum f\"ur Luft- und Raumfahrt e. V. (DLR), Linder H\"ohe, 51147 K\"oln, Germany\label{aff31}
\and
INAF-Osservatorio Astronomico di Roma, Via Frascati 33, 00078 Monteporzio Catone, Italy\label{aff32}
\and
INFN section of Naples, Via Cinthia 6, 80126, Napoli, Italy\label{aff33}
\and
Institute for Astronomy, University of Hawaii, 2680 Woodlawn Drive, Honolulu, HI 96822, USA\label{aff34}
\and
Dipartimento di Fisica e Astronomia "Augusto Righi" - Alma Mater Studiorum Universit\`a di Bologna, Viale Berti Pichat 6/2, 40127 Bologna, Italy\label{aff35}
\and
Instituto de Astrof\'{\i}sica de Canarias, E-38205 La Laguna, Tenerife, Spain\label{aff36}
\and
Institute for Astronomy, University of Edinburgh, Royal Observatory, Blackford Hill, Edinburgh EH9 3HJ, UK\label{aff37}
\and
Jodrell Bank Centre for Astrophysics, Department of Physics and Astronomy, University of Manchester, Oxford Road, Manchester M13 9PL, UK\label{aff38}
\and
European Space Agency/ESRIN, Largo Galileo Galilei 1, 00044 Frascati, Roma, Italy\label{aff39}
\and
Universit\'e Claude Bernard Lyon 1, CNRS/IN2P3, IP2I Lyon, UMR 5822, Villeurbanne, F-69100, France\label{aff40}
\and
Institut de Ci\`{e}ncies del Cosmos (ICCUB), Universitat de Barcelona (IEEC-UB), Mart\'{i} i Franqu\`{e}s 1, 08028 Barcelona, Spain\label{aff41}
\and
Instituci\'o Catalana de Recerca i Estudis Avan\c{c}ats (ICREA), Passeig de Llu\'{\i}s Companys 23, 08010 Barcelona, Spain\label{aff42}
\and
Institut de Ciencies de l'Espai (IEEC-CSIC), Campus UAB, Carrer de Can Magrans, s/n Cerdanyola del Vall\'es, 08193 Barcelona, Spain\label{aff43}
\and
UCB Lyon 1, CNRS/IN2P3, IUF, IP2I Lyon, 4 rue Enrico Fermi, 69622 Villeurbanne, France\label{aff44}
\and
Mullard Space Science Laboratory, University College London, Holmbury St Mary, Dorking, Surrey RH5 6NT, UK\label{aff45}
\and
Departamento de F\'isica, Faculdade de Ci\^encias, Universidade de Lisboa, Edif\'icio C8, Campo Grande, PT1749-016 Lisboa, Portugal\label{aff46}
\and
Instituto de Astrof\'isica e Ci\^encias do Espa\c{c}o, Faculdade de Ci\^encias, Universidade de Lisboa, Campo Grande, 1749-016 Lisboa, Portugal\label{aff47}
\and
Department of Astronomy, University of Geneva, ch. d'Ecogia 16, 1290 Versoix, Switzerland\label{aff48}
\and
Universit\'e Paris-Saclay, CNRS, Institut d'astrophysique spatiale, 91405, Orsay, France\label{aff49}
\and
INFN-Padova, Via Marzolo 8, 35131 Padova, Italy\label{aff50}
\and
Aix-Marseille Universit\'e, CNRS/IN2P3, CPPM, Marseille, France\label{aff51}
\and
Max Planck Institute for Extraterrestrial Physics, Giessenbachstr. 1, 85748 Garching, Germany\label{aff52}
\and
Universit\"ats-Sternwarte M\"unchen, Fakult\"at f\"ur Physik, Ludwig-Maximilians-Universit\"at M\"unchen, Scheinerstr.~1, 81679 M\"unchen, Germany\label{aff53}
\and
INAF-Istituto di Astrofisica e Planetologia Spaziali, via del Fosso del Cavaliere, 100, 00100 Roma, Italy\label{aff54}
\and
Space Science Data Center, Italian Space Agency, via del Politecnico snc, 00133 Roma, Italy\label{aff55}
\and
INFN-Bologna, Via Irnerio 46, 40126 Bologna, Italy\label{aff56}
\and
University Observatory, LMU Faculty of Physics, Scheinerstr.~1, 81679 Munich, Germany\label{aff57}
\and
Institute of Theoretical Astrophysics, University of Oslo, P.O. Box 1029 Blindern, 0315 Oslo, Norway\label{aff58}
\and
Felix Hormuth Engineering, Goethestr. 17, 69181 Leimen, Germany\label{aff59}
\and
Technical University of Denmark, Elektrovej 327, 2800 Kgs. Lyngby, Denmark\label{aff60}
\and
Cosmic Dawn Center (DAWN), Denmark\label{aff61}
\and
Max-Planck-Institut f\"ur Astronomie, K\"onigstuhl 17, 69117 Heidelberg, Germany\label{aff62}
\and
NASA Goddard Space Flight Center, Greenbelt, MD 20771, USA\label{aff63}
\and
Department of Physics and Astronomy, University College London, Gower Street, London WC1E 6BT, UK\label{aff64}
\and
Department of Physics and Helsinki Institute of Physics, Gustaf H\"allstr\"omin katu 2, University of Helsinki, 00014 Helsinki, Finland\label{aff65}
\and
Universit\'e Paris-Saclay, Universit\'e Paris Cit\'e, CEA, CNRS, AIM, 91191, Gif-sur-Yvette, France\label{aff66}
\and
Universit\'e de Gen\`eve, D\'epartement de Physique Th\'eorique and Centre for Astroparticle Physics, 24 quai Ernest-Ansermet, CH-1211 Gen\`eve 4, Switzerland\label{aff67}
\and
Department of Physics, P.O. Box 64, University of Helsinki, 00014 Helsinki, Finland\label{aff68}
\and
Helsinki Institute of Physics, Gustaf H{\"a}llstr{\"o}min katu 2, University of Helsinki, 00014 Helsinki, Finland\label{aff69}
\and
Laboratoire d'\'etude de l'Univers et des ph\'enom\`enes eXtr\^emes, Observatoire de Paris, Universit\'e PSL, Sorbonne Universit\'e, CNRS, 92190 Meudon, France\label{aff70}
\and
Aix-Marseille Universit\'e, CNRS, CNES, LAM, Marseille, France\label{aff71}
\and
SKAO, Jodrell Bank, Lower Withington, Macclesfield SK11 9FT, UK\label{aff72}
\and
Centre de Calcul de l'IN2P3/CNRS, 21 avenue Pierre de Coubertin 69627 Villeurbanne Cedex, France\label{aff73}
\and
Dipartimento di Fisica "Aldo Pontremoli", Universit\`a degli Studi di Milano, Via Celoria 16, 20133 Milano, Italy\label{aff74}
\and
INFN-Sezione di Milano, Via Celoria 16, 20133 Milano, Italy\label{aff75}
\and
Universit\"at Bonn, Argelander-Institut f\"ur Astronomie, Auf dem H\"ugel 71, 53121 Bonn, Germany\label{aff76}
\and
INFN-Sezione di Roma, Piazzale Aldo Moro, 2 - c/o Dipartimento di Fisica, Edificio G. Marconi, 00185 Roma, Italy\label{aff77}
\and
Dipartimento di Fisica e Astronomia "Augusto Righi" - Alma Mater Studiorum Universit\`a di Bologna, via Piero Gobetti 93/2, 40129 Bologna, Italy\label{aff78}
\and
Department of Physics, Institute for Computational Cosmology, Durham University, South Road, Durham, DH1 3LE, UK\label{aff79}
\and
Universit\'e C\^{o}te d'Azur, Observatoire de la C\^{o}te d'Azur, CNRS, Laboratoire Lagrange, Bd de l'Observatoire, CS 34229, 06304 Nice cedex 4, France\label{aff80}
\and
Universit\'e Paris Cit\'e, CNRS, Astroparticule et Cosmologie, 75013 Paris, France\label{aff81}
\and
CNRS-UCB International Research Laboratory, Centre Pierre Bin\'etruy, IRL2007, CPB-IN2P3, Berkeley, USA\label{aff82}
\and
Institut d'Astrophysique de Paris, 98bis Boulevard Arago, 75014, Paris, France\label{aff83}
\and
Institut d'Astrophysique de Paris, UMR 7095, CNRS, and Sorbonne Universit\'e, 98 bis boulevard Arago, 75014 Paris, France\label{aff84}
\and
Institute of Physics, Laboratory of Astrophysics, Ecole Polytechnique F\'ed\'erale de Lausanne (EPFL), Observatoire de Sauverny, 1290 Versoix, Switzerland\label{aff85}
\and
Telespazio UK S.L. for European Space Agency (ESA), Camino bajo del Castillo, s/n, Urbanizacion Villafranca del Castillo, Villanueva de la Ca\~nada, 28692 Madrid, Spain\label{aff86}
\and
Institut de F\'{i}sica d'Altes Energies (IFAE), The Barcelona Institute of Science and Technology, Campus UAB, 08193 Bellaterra (Barcelona), Spain\label{aff87}
\and
DARK, Niels Bohr Institute, University of Copenhagen, Jagtvej 155, 2200 Copenhagen, Denmark\label{aff88}
\and
Waterloo Centre for Astrophysics, University of Waterloo, Waterloo, Ontario N2L 3G1, Canada\label{aff89}
\and
Department of Physics and Astronomy, University of Waterloo, Waterloo, Ontario N2L 3G1, Canada\label{aff90}
\and
Perimeter Institute for Theoretical Physics, Waterloo, Ontario N2L 2Y5, Canada\label{aff91}
\and
Centre National d'Etudes Spatiales -- Centre spatial de Toulouse, 18 avenue Edouard Belin, 31401 Toulouse Cedex 9, France\label{aff92}
\and
Institute of Space Science, Str. Atomistilor, nr. 409 M\u{a}gurele, Ilfov, 077125, Romania\label{aff93}
\and
Dipartimento di Fisica e Astronomia "G. Galilei", Universit\`a di Padova, Via Marzolo 8, 35131 Padova, Italy\label{aff94}
\and
Institut f\"ur Theoretische Physik, University of Heidelberg, Philosophenweg 16, 69120 Heidelberg, Germany\label{aff95}
\and
Institut de Recherche en Astrophysique et Plan\'etologie (IRAP), Universit\'e de Toulouse, CNRS, UPS, CNES, 14 Av. Edouard Belin, 31400 Toulouse, France\label{aff96}
\and
Universit\'e St Joseph; Faculty of Sciences, Beirut, Lebanon\label{aff97}
\and
Departamento de F\'isica, FCFM, Universidad de Chile, Blanco Encalada 2008, Santiago, Chile\label{aff98}
\and
Universit\"at Innsbruck, Institut f\"ur Astro- und Teilchenphysik, Technikerstr. 25/8, 6020 Innsbruck, Austria\label{aff99}
\and
Institut d'Estudis Espacials de Catalunya (IEEC),  Edifici RDIT, Campus UPC, 08860 Castelldefels, Barcelona, Spain\label{aff100}
\and
Satlantis, University Science Park, Sede Bld 48940, Leioa-Bilbao, Spain\label{aff101}
\and
Institute of Space Sciences (ICE, CSIC), Campus UAB, Carrer de Can Magrans, s/n, 08193 Barcelona, Spain\label{aff102}
\and
Infrared Processing and Analysis Center, California Institute of Technology, Pasadena, CA 91125, USA\label{aff103}
\and
Instituto de Astrof\'isica e Ci\^encias do Espa\c{c}o, Faculdade de Ci\^encias, Universidade de Lisboa, Tapada da Ajuda, 1349-018 Lisboa, Portugal\label{aff104}
\and
Cosmic Dawn Center (DAWN)\label{aff105}
\and
Niels Bohr Institute, University of Copenhagen, Jagtvej 128, 2200 Copenhagen, Denmark\label{aff106}
\and
Universidad Polit\'ecnica de Cartagena, Departamento de Electr\'onica y Tecnolog\'ia de Computadoras,  Plaza del Hospital 1, 30202 Cartagena, Spain\label{aff107}
\and
Caltech/IPAC, 1200 E. California Blvd., Pasadena, CA 91125, USA\label{aff108}
\and
Instituto de F\'isica Te\'orica UAM-CSIC, Campus de Cantoblanco, 28049 Madrid, Spain\label{aff109}
\and
Zentrum f\"ur Astronomie, Universit\"at Heidelberg, Philosophenweg 12, 69120 Heidelberg, Germany\label{aff110}
\and
ICL, Junia, Universit\'e Catholique de Lille, LITL, 59000 Lille, France\label{aff111}}    

\abstract{We present an example cluster follow-up study with \textit{Euclid}. Our target, a $z\sim 1.74$ candidate cluster nicknamed the `Puddle', was initially discovered by the Massive and Distant Clusters of WISE Survey 2 as a $z_\mathrm{phot}\sim 1.65$ candidate cluster. It was also detected independently as a $z_\mathrm{phot}\sim 1.5$ candidate with the two cluster-finding algorithms in Euclid Quick Release 1 (Q1). A Keck MOSFIRE spectrum shows the brightest nucleus is at $z=1.74$ and is dominated by an active galactic nucleus. Our analysis focused on the galaxy population and the brightest cluster galaxy (BCG), and is based on \textit{Euclid} and ancillary photometry. Compared to similar fields, we measured an overdensity of $110\pm 14$ galaxies with $\HE\leq 22.25$ in a $\ang{;2}$ radius around the BCG. About $18\pm 4$\% of the completeness-corrected galaxy population is red, which is consistent with some clusters at $z>1.5$ but lower than others. \textit{Euclid} imaging revealed that six or seven galaxies appear to be assembling to form the future BCG. Spectral energy distribution fitting suggests that the merging BCG has a stellar mass of $5.7\pm 0.3\times 10^{11}\,M_\odot$ and that it experienced a short burst of star formation $\sim 300\,$Myr ago. Its morphology and star-formation history suggest that the proto-BCG is a more evolved version of the merging core of SPT2349$-$56. These systems indicate that multiobject mergers might be a common BCG formation process. Assuming a similar density of mergers in the Euclid Wide Survey, we expect that \textit{Euclid} will discover approximately 400 assembling BCGs by the end of its mission.}

\keywords{Galaxies: active, Galaxies: clusters: individual: MOO2$\,$J03374$-$28386/EUCL$-$Q1$-$CL$\,$J033730.18$-$283827.6, Galaxies: formation, Galaxies: interactions, Galaxies: star formation}

\titlerunning{\Euclid\/: A blue galaxy population and a BCG in the making in a $z\sim 1.74$ galaxy cluster candidate}
\authorrunning{A.~Trudeau et al.}

\maketitle

\section{Introduction}\label{sec_intro}

Galaxy clusters sit at the nodes of the cosmic web and are linked by filaments \citep{bond_how_1996,springel_simulations_2005,vogelsberger_introducing_2014}. As the maxima of the density field in the Universe, galaxy clusters are natural laboratories to test a variety of astrophysical phenomena in different fields: cosmology \citep[e.g.][]{vikhlinin_chandra_2009,allen_cosmological_2011,pierre_xxl_2016}, plasma and high-energy physics \citep[e.g.][]{mcnamara_heating_2007,mcnamara_mechanical_2012,zhuravleva_turbulent_2014},
and galaxy evolution \citep[e.g.][]{von_der_linden_star_2010,peng_mass_2010,peng_mass_2012}.

The cessation of star formation in galaxies, usually called `quenching', depends on two main factors: galaxy mass \citep{kauffmann_dependence_2003} and environment density \citep{peng_mass_2010,peng_mass_2012}. Dense environments such as local galaxy clusters tend to host a larger fraction of quiescent galaxies than field galaxy samples \citep[e.g.][]{dressler_galaxy_1980,poggianti_star_1999,balogh_bimodal_2004,balogh_colour_2009,balogh_evidence_2016,kawinwanichakij_effect_2017,pintos-castro_evolution_2019,ragusa_galaxies_2025}. At $z<1$ the fraction of quenched galaxies in clusters decreases with increasing redshift \citep[e.g.][]{balogh_bimodal_2004,raichoor_galaxy_2012,pintos-castro_evolution_2019} but always remains above the field level. The situation is less clear at $z\gtrsim 1.5$. Some authors find that high-redshift galaxy clusters already exhibit elevated levels of quenching \citep[e.g.][]{newman_spectroscopic_2014,cooke_submillimetre_2019,lemaux_persistence_2019,strazzullo_galaxy_2019,van_der_burg_gogreen_2020,toni_cosmos-web_2026}, while others observe no difference with field levels before $z\sim 1.5$ \citep[e.g.][]{tran_reversal_2010,brodwin_era_2013,alberts_star_2016,nantais_halpha_2020,trudeau_massive_2024}.

Brightest cluster galaxies (BCGs) represent another example of the interplay between galaxy evolution and cluster environment. BCGs are the most massive galaxies in the Universe, and they are found close to the centre of the potential wells of galaxy clusters \citep{zitrin_miscentring_2012,hashimoto_multiwavelength_2014,cui_how_2016,lopes_optical_2018}. Their properties (e.g. shape, mass, star-formation history) tend to correlate with those of their host clusters \citep[e.g.][]{ebeling_extreme_2021}. A particular example is the relation between the active galactic nucleus (AGN) activity in BCGs and the regulation of the temperature of the intracluster medium \citep[ICM;][]{hu_long-slit_1985,burns_radio_1990,cavagnolo_entropy_2008,rafferty_regulation_2008,hlavacek-larrondo_extreme_2012}. An imbalance between the cooling of the ICM and the energy injected by the AGN \citep[e.g.][]{mcnamara_star_1989,fabian_cooling_1994,mcnamara_heating_2005,rafferty_feedback-regulated_2006} can trigger gas condensation onto the BCG, a phenomenon called a `cooling flow'.

 The BCG evolution is intimately linked to the evolution of its host cluster. At $z\lesssim 1$, BCG evolution is dominated by gas-poor mergers \citep[e.g.][]{aragon-salamanca_k-band_1998,dubinski_origin_1998,lidman_evidence_2012,lidman_importance_2013,burke_growth_2013,golden-marx_impact_2018,golden-marx_hierarchical_2025,montenegro-taborda_growth_2023}. While most local BCGs are quiescent, between 20\% and 35\% of them exhibit small amounts of star formation \citep[usually below the field level, see][]{orellana-gonzalez_evolution_2022} powered by residual cooling flows \citep[e.g.][]{crawford_rosat_1999,rawle_relation_2012,oliva-altamirano_galaxy_2014,mcdonald_star-forming_2016}. 

However, the formation and early evolution of BCGs remain poorly understood. A popular formation model \citep{de_lucia_hierarchical_2007} predicted that 80\% of the stellar mass of $z=0$ BCGs would form before $z\sim 3$ in separate progenitors that progressively assemble into a BCG through gas-poor mergers. Observations of $z\gtrsim 1$ BCGs have shown a different picture: high-redshift BCGs are diverse, but some of them display substantial levels of in situ star formation \citep{webb_star_2015,mcdonald_star-forming_2016,bonaventura_red_2017}. The cause of the star formation is not currently understood. Some authors \citep[e.g.][]{rennehan_rapid_2020,remus_young_2023} suggest that it might be powered by the initial formation of the BCG in a gas-rich merger involving multiple galaxies. Observations of protocluster cores with elevated star-formation rates (SFRs) and interacting galaxies support this scenario \citep{miley_spiderweb_2006,kuiper_sinfoni_2011,miller_massive_2018,coogan_z_2023}. One cluster core at $z=1.71$, SpARCS104922.6+564032.5 \citep{webb_extreme_2015,hlavacek-larrondo_evidence_2020} indicates that massive cooling flows could also power star formation.

Our current understanding of quenching in high-redshift clusters and BCG early evolution are limited by small number statistics since most cluster samples contain only a few objects at $z\gtrsim 1.5$ \citep[e.g.][]{wilson_clusters_2006,wilson_spectroscopic_2009,wylezalek_galaxy_2013,bleem_galaxy_2015,pierre_xxl_2016,bulbul_erosita_2022,bulbul_srgerosita_2024}. There are several recent protocluster searches that explore these redshifts \citep[e.g.][]{martinache_spitzer_2018,ouchi_systematic_2018,toshikawa_goldrush_2018,golden-marx_high-redshift_2019,ando_systematic_2020,gully_spitzer-selected_2024}, but their sample purities tend to be low \citep[e.g.][]{hung_discovering_2025}.

With the advent of a new generation of large-scale survey facilities \citep[\textit{Euclid}, \textit{Nancy Grace Roman}, \textit{Vera C. Rubin,} see][]{spergel_wide-field_2015,ivezic_lsst_2019,robertson_galaxy_2019,Scaramella-EP1,grishin_yolo-cl_2025}, this situation is changing, and large samples of (proto)clusters are becoming available. \textit{Euclid} is the only one of these facilities that is currently in service. Located at the second Sun-Earth Lagrange point, \textit{Euclid} has a 1.2-metre primary mirror \citep{EuclidSkyOverview}. By the end of its six-year mission, \textit{Euclid} is expected to observe $14\,000\,$deg$^{2}$ of the sky \citep{EuclidSkyOverview} and to find about $3 \times 10^5$ galaxy clusters at $1<z<2$, according to the latest estimates \citep[][see also the estimations of \citealt{Adam-EP3} and \citealt{sartoris_next_2016}]{EuclidSkyOverview}.

 The \textit{Euclid} search for galaxy clusters uses two algorithms: the Adaptive Matched Identifier of Clustered Objects (AMICO) and PZWav \citep[][]{Adam-EP3,Q1-SP050}. AMICO is a matched filter algorithm and it detects clusters with a redshift-dependent filter mimicking the sum of the cluster contribution and the noise \citep{bellagamba_amico_2018,maturi_amico_2019}. PZWav instead applies a spatial filter constructed from the difference of two Gaussians. The width of the Gaussians was chosen to select overdensities with sizes consistent with a cluster extent on the sky \citep{thongkham_massive_2024a}.

 The PZWav algorithm is also used by another large-scale cluster survey, the Massive and Distant Clusters of WISE Survey 2 \citep[MaDCoWS2;][]{thongkham_massive_2024a,thongkham_massive_2024b}. MaDCoWS2 derives photometric redshifts \citep{brodwin_photometric_2006} using data from the Dark Energy Camera (DECam) and the Wide-Field Infrared Survey Explorer (WISE), and it uses these photometric redshifts as input to PZWav. While the MaDCoWS2 survey identified 6959 candidate clusters at $z_\mathrm{phot}\geq 1.5$, the resolution of WISE -- the point spread functions (PSFs) full widths at half maxima (FWHM) are $\ang{;;6.1}$ and $\ang{;;6.4}$ in the two bluest channels \citep[see][]{wright_wide-field_2010} -- is insufficient for distinguishing individual galaxies in the packed cores of high-redshift clusters. Near-infrared data sets with better resolution, such as \textit{Euclid} images, are needed to enable galaxy evolution studies of MaDCoWS2 clusters. Furthermore, the Euclid Wide Survey will largely overlap with the MaDCoWS2 footprint \citep{Scaramella-EP1,EuclidSkyOverview}.

\begin{figure*}[htbp!]
\centering
\includegraphics[angle=0,width=0.49\textwidth]{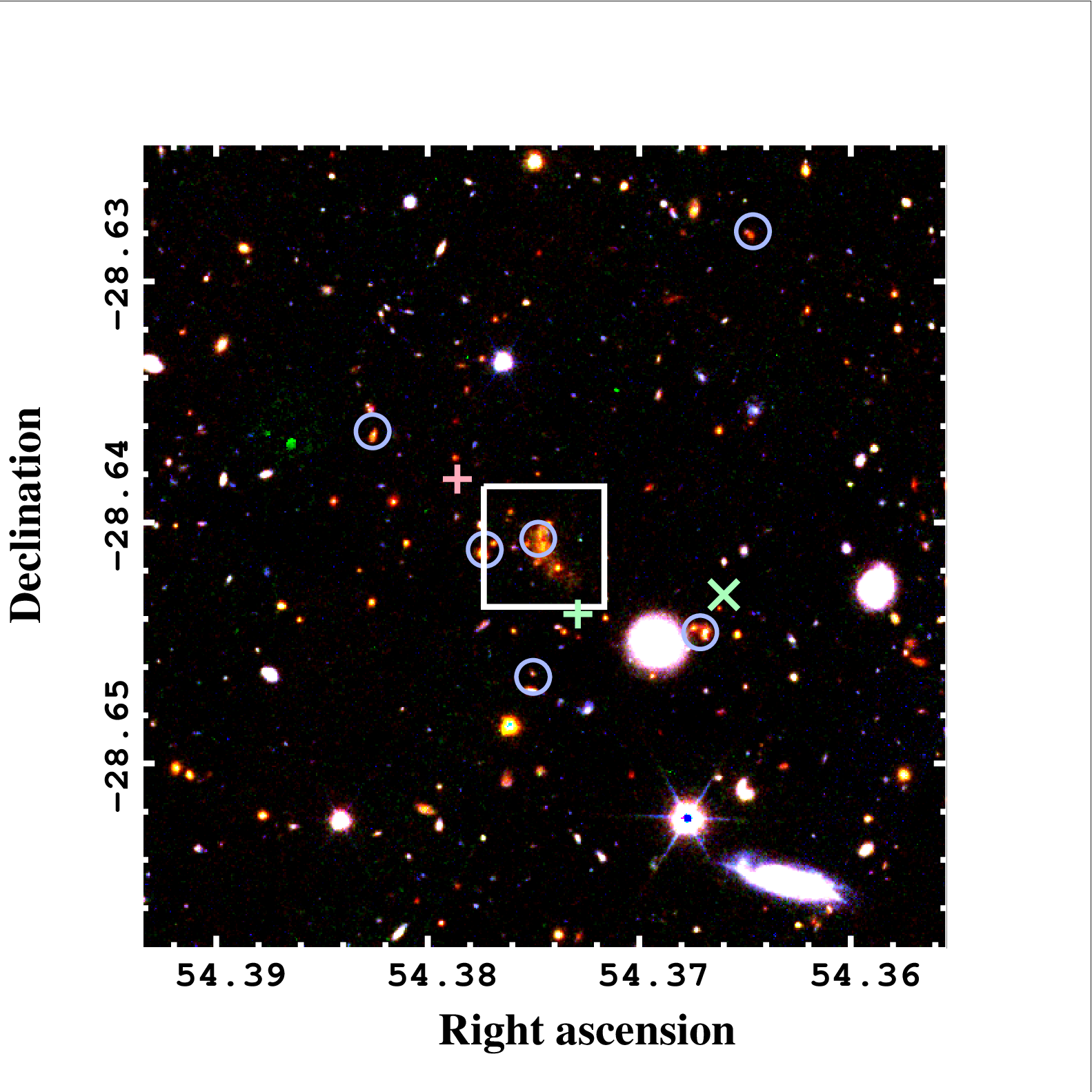}
\includegraphics[angle=0,width=0.49\textwidth]{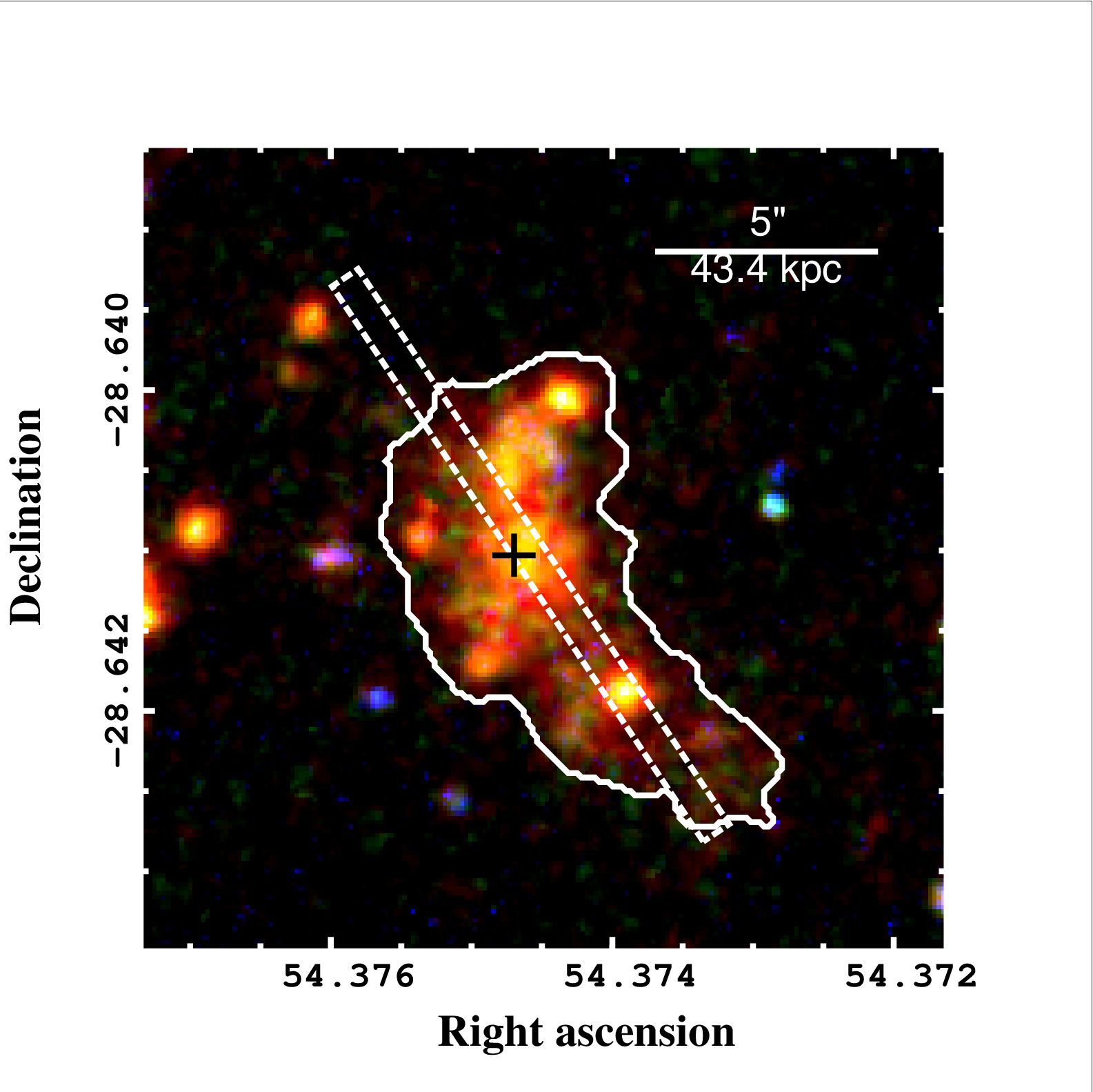}
\caption{\emph{Left}: Tricolour image (\IE, \YE, and \HE) of the Puddle cluster with a $\ang{;2} \times \ang{;2}$ field of view. The position of the MaDCoWS2 detection is indicated by a light green `X'. The coordinates of the Puddle cluster in \citet{Q1-SP050} are indicated by crosses: light green for the PZWav coordinates and pink for the AMICO coordinates. The most likely photometric members of MaDCoWS2 (galaxies for which the integrated probability density function between $z_\mathrm{phot}-\sigma_z$ and $z_\mathrm{phot}+\sigma_z$ is equal or greater than 0.3) are circled in mauve. The white box indicates the $\ang{;0.3} \times \ang{;0.3}$ field of view of the zoom-in in the right panel. \emph{Right}: View of the BCG complex in the Puddle cluster. The white contours indicate the edges of the aperture used to measure the photometry. The white dashed box shows the approximate position of the spectroscopic slit. The centre of the MIPS $24\,\mathrm{\micron}$ emission is indicated by a black cross.
}
\label{fig_presentation}
\end{figure*}

In this paper, we present \textit{Euclid}'s first view of a candidate cluster detected in both MaDCoWS2 and the \textit{Euclid} preliminary cluster search: MOO2$\,$J03374$-$28386/EUCL$-$Q1$-$CL$\,$J033730.18$-$283827.6. We nicknamed this candidate the `Puddle' cluster due to the intriguing BCG complex revealed by \textit{Euclid}. The left panel of Fig. \ref{fig_presentation} presents a $\ang{;2} \times \ang{;2}$ tricolour image based on \textit{Euclid} data; the right panel shows a zoomed $\ang{;0.3} \times \ang{;0.3}$ view of the BCG complex. This system represents a test of the capabilities -- and limitations -- of \textit{Euclid} for the discovery and follow-up of high-redshift galaxy clusters. 

The paper is divided as follows: Sect. \ref{sec_data} presents the data used to discover and characterise the Puddle cluster. Section \ref{sec_analysis} presents the data analysis, focusing on the galaxy population and the central structure (i.e. the BCG complex) characterisation. In Sect. \ref{sec_discussion} we discuss our findings, which are summarised in Sect. \ref{sec_conclusion}. Throughout this paper, we assume a \citet{collaboration_planck_2020} $\Lambda$ cold dark matter ($\Lambda$CDM) cosmology as implemented in \texttt{astropy.cosmology}: $\Omega\mathrm{_m} = 0.31$ and $H_0 = 67.7\,\kmsMpc$. At $1.5<z<2.0$, an angular distance of \ang{;2} corresponds to a proper distance between 1.03 and 1.04$\,$Mpc. All magnitudes are in the AB system. We assumed a \citet{kroupa_mass_2002} initial mass function.

\section{Data}\label{sec_data}

\subsection{MaDCoWS2 catalogue}\label{ssec_madcows2}
MaDCoWS2 is an optical and near-infrared catalogue that covers $6498\,\mathrm{deg^2}$ and contains 133\,036 candidate galaxy clusters \citep{thongkham_massive_2024a,thongkham_massive_2024b}. It is based on the $grz$ bands from the DECam Legacy Survey \citep[DECaLS; see][]{flaugher_dark_2015,dey_overview_2019} and the $W1W2$ bands from the CatWISE2020 catalogue \citep{eisenhardt_catwise_2020,marocco_catwise2020_2021}. The photometric redshift probability distribution functions \citep[PDFs; see][]{brodwin_photometric_2006} computed from these bands was input in PZWav. In MaDCoWS2, the Puddle cluster was detected as a $z_\mathrm{phot}=1.65\pm 0.08$ overdensity with a signal-to-noise ratio (S\slash N) of 5.6. 

\subsection{Euclid Q1 data release}\label{ssec_Q1}

\textit{Euclid}'s two instruments are the Visible Camera \citep[VIS;][]{EuclidSkyVIS}, a single-band (denoted \IE) imaging camera covering the $5400$ to $9200\,$\AA~range, and the Near-Infrared Spectrometer and Photometer \citep[NISP;][]{EuclidSkyNISP}. The NISP can perform imaging (\YE, \JE, and \HE bands) and multiobject slitless spectroscopy (blue and red grisms).

The Euclid Consortium Quick Release 1 \citep[hereafter the Q1 release,][]{Q1cite,EuclidSkyOverview,Q1-TP001} is composed primarily of \textit{Euclid} images and photometric and spectroscopic catalogues \citep[MER and SPE catalogues, see][]{Q1-TP001,Q1-TP004,Q1-TP006} of the Euclid Deep Field North, Euclid Deep Field Fornax, and Euclid Deep Field South. The two last fields are also covered by the MaDCoWS2 catalogue. The Puddle cluster is located in the Euclid Deep Field Fornax.

A preliminary cluster search is also part of the Q1 release \citep{Q1-SP050}. This search focuses on $0.2 \lesssim z\lesssim 1.5$ detections. \textit{Euclid} photometric redshifts are still being refined and currently have large uncertainties at $z\gtrsim 1.5$. However, \citet{Q1-SP050} includes a sample of 15 unconfirmed $z\gtrsim 1.5$ clusters that meet the requirement of being detected both by PZWav and AMICO. The Puddle cluster is among this list. It was detected by PZWav as a $z_\mathrm{phot}=1.51$ overdensity with an S\slash N of 9.17 and by AMICO as a $z_\mathrm{phot}=1.49$ overdensity with an S\slash N of 16.39 and a richness of $12.05\pm 1.88$. AMICO and PZWAv compute S\slash N differently; hence AMICO produces a systematically higher S\slash N than PZWAv (see Fig. 4 of \citealt{Q1-SP050}). However, no redshift uncertainties are provided, and the scatter between photometric and spectroscopic redshifts is estimated for $z<1.5$ only. We therefore used the MaDCoWS2 detection as our primary reference for the cluster position and photometric redshift.

\subsubsection{MER catalogue}\label{sssec_catalog}

The Q1 MER catalogue\footnote{available in Euclid Science Archive System, at \url{https://eas.esac.esa.int/sas/}} combines VIS, NISP, and several external data sets, including $griz$ bands from DECam \citep[][]{honscheid_dark_2008,flaugher_dark_2015} on the 4$\,$m Blanco Telescope, to create a common catalogue with consistently measured photometry. \textit{Euclid} and external images are organised in calibrated $\ang{;32} \times \ang{;32}$ tiles with $\ang{;2}$ overlaps and $\ang{;;0.1}$ pixel scale \citep{Q1-TP004}. Sources are detected in VIS and NISP, de-blended, and organised into a single source list which is then used to measure photometry in all bands.

To identify the presence of a red sequence, we used the $z$ and \HE bands, which bracket the $4000\,$\AA~break at $1.5 \lesssim z \lesssim 2.75$. While the \IE band lies blueward of the $4000\,$\AA~break at these redshifts, its broadness \citep[effective width of $3318.32\,$\AA, with a mean wavelength of $7334.36\,$\AA, see][]{Rodrigo_SVO_2012,Rodrigo_SVO_2020,rodrigo_photometric_2024} limits its usefulness for characterising galaxy populations. Specifically, we used the $z$ and \HE fluxes obtained via a single S\'ersic model-fitting, denoted by a `\texttt{\_sersic}' suffix in the MER catalogue.

\subsubsection{MER images}\label{sssec_euclid_images}

The MER catalogue pipeline performs well for most galaxies, but tends to overcorrect or suppress low-surface brightness features like tidal tails \citep{Q1-TP001,Q1-SP003}. Furthermore, the single S\'ersic and circular aperture photometry currently available are poorly suited to irregular galaxies. Thus, rather than rely on the MER catalogue photometry of the BCG complex, we downloaded \textit{Euclid} and DECam level 2 mosaics of the Puddle cluster, and trimmed them to a more manageable size of $\ang{;6} \times \ang{;6}$. We estimated the flux of the BCG with \texttt{sep} \citep{barbary_sep_2018}, using a custom aperture (see Sect. \ref{ssec_bcg}). The level 2 DECam mosaics in the Euclid Science Archive System are already resampled to the pixel scale of the \textit{Euclid} images. 

\subsection{\textit{Spitzer} data}\label{ssec_spitzer}

The \textit{Spitzer} Space Telescope \citep{werner_spitzer_2004} was a near- and mid-infrared observatory launched in August 2003 with three instruments on board: the Infrared Array Camera \citep[IRAC,][]{fazio_infrared_2004}, the Multiband Infrared Photometer for \textit{Spitzer} \citep[MIPS; ][]{rieke_multiband_2004}, and the Infrared Spectrograph \citep[IRS,][]{houck_infrared_2004}. IRAC had four bands, usually called `channels', centred on $3.6$, $4.5$, $5.8$, and $8.0\,\mathrm{\micron}$; MIPS had three bands, centred on $24$, $70$, and $160\,\mathrm{\micron}$. However, after \textit{Spitzer} exhausted its liquid helium reserve in May 2009, only the two bluest bands of the IRAC camera could be used.

The Cosmic Dawn Survey \citep[][see also \citealt{EP-Zalesky}]{Moneti-EP17} is an IRAC survey covering the three Euclid Deep Fields and several calibration fields. Its image mosaics are based on both dedicated and archival data in the four IRAC channels, and correspond to about 11\% of the telescope total mission time \citep{Moneti-EP17}. A source catalogue is available; however, the aperture sizes are inappropriate for the shape of the BCG. Furthermore, the BCG complex is blended
with many neighbouring galaxies at $3.6$ and $4.5\,\mathrm{\micron}$ and is barely detected at $5.8$ and $8.0\,\micron$. We thus estimated the flux directly from the mosaics (see Sect. \ref{ssec_bcg}). 

The coordinates of the brightest core in the BCG complex correspond ($\lesssim \ang{;;0.6}$ separation) to a point source detected at $24\,\mathrm{\micron}$ in the Spitzer Wide-area InfraRed Extragalactic Survey \citep[SWIRE; see][]{lonsdale_swire:_2003,lonsdale_first_2004}. Given the relative isolation of this source (the closest MIPS source is $\ang{;;26}$ away) and the $\ang{;;6}$ FHWM of the MIPS $24\,\mathrm{\micron}$ point-spread function \citep[PSF; ][]{rieke_multiband_2004}, we adopted the SWIRE $24\,\mathrm{\micron}$ flux here.

\subsection{\textit{Keck} MOSFIRE spectroscopy}\label{ssec_spect_obs}

The Euclid Q1 release contains NISP slitless multiobject spectra taken with the red grism \citep{Q1-TP006,EuclidSkyNISP}. However, the reduction of the grism data was not optimal in Q1; most of the identified issues are being addressed for the Euclid DR1. In the Puddle cluster field of view, an average of 55.6\% of the pixels of the calibrated unidimensional spectra (excluding the edges) are flagged for suspicious behaviour.

Thus, to determine the Puddle cluster redshift, we obtained \textit{Keck} MOSFIRE multiobject spectroscopy in the $H$ band. The observations were taken on the night of 24-25 December 2024, with a total on-target integration time of $5726\,$s. The weather was clear with an airmass between 1.5 and 1.7 during the observations, and the average seeing was about \ang{;;0.7}. We used $119\,$s integrations on each node of a 2-point dither pattern with a $\ang{;;3.6}$ separation between the nodes, and a single mask with 18 slits (usually $\mathrm{\ang{;;14.99}\times \ang{;;0.7}}$) for the whole observation. The targeted objects were selected to have magnitudes consistent with high-redshift galaxies in the MER catalogue and different $z-\HE$ colours. One of the slits was positioned across the BCG complex; its 1 and 2D spectra are presented in Sect. \ref{ssec_spectro}.

\begin{figure}[ht!]
\centering
\includegraphics[angle=0,width=\columnwidth]{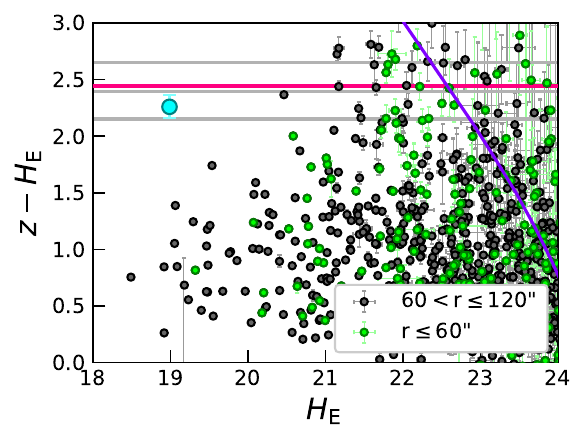}
\caption{$z-\HE$ CMD for a $\ang{;2}$ region centred on the BCG. Only the galaxies with S\slash N$>3\,\sigma$ in $z$ and \HE bands are shown. The expected location of the red sequence at $z = 1.74$ is shown in pink, while the red sequences at $z = 1.4$, $z = 1.7$, and $z = 2.0$ are in grey. The photometry of the BCG complex is indicated by a cyan symbol. The solid purple line indicates the 50\% combined $z$ and \HE completeness limit.}
\label{fig_cmd}
\end{figure}

\begin{figure*}[ht!]
\centering
\includegraphics[angle=0,width=\textwidth]{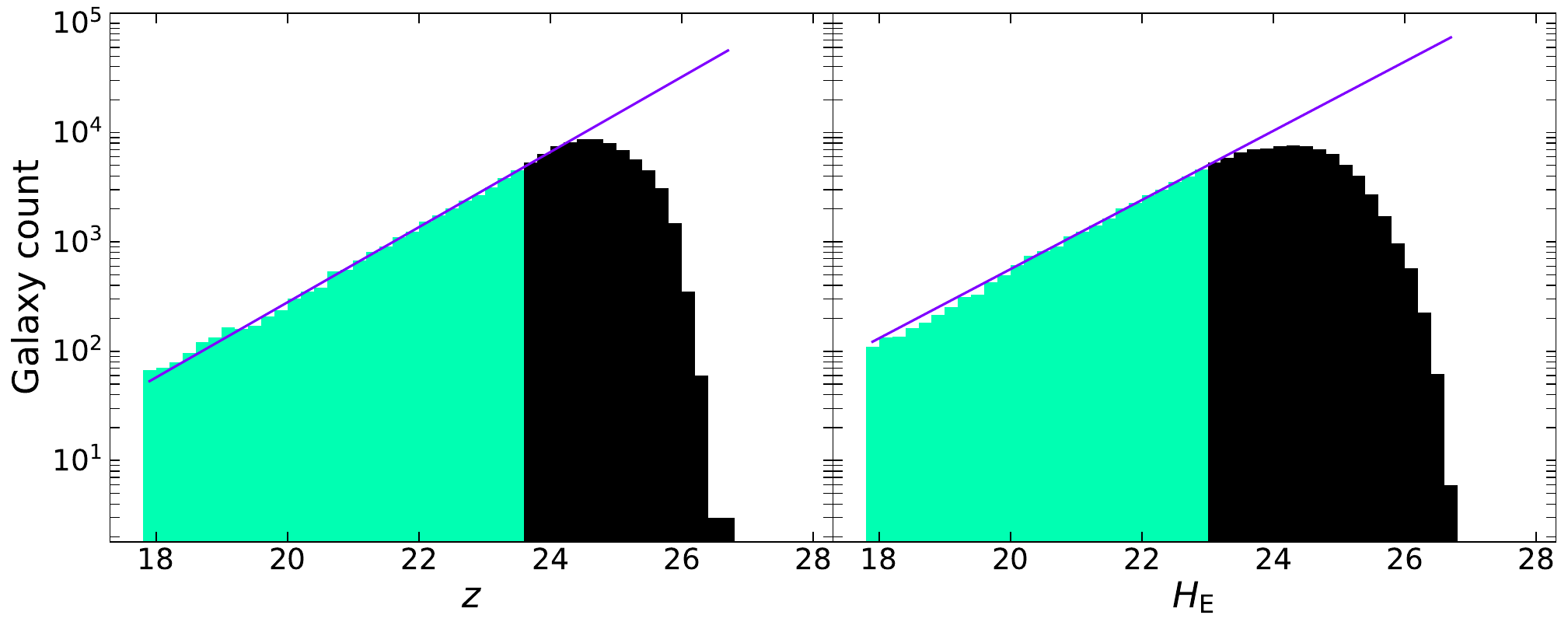}
\caption{\emph{Left}: Computation of the completeness correction for the $z$ band. The galaxy counts in the field are in green in the region where it is considered complete and black elsewhere. The purple line indicates the fit used to compute the completeness correction. \emph{Right}: Same but for the \HE band.}
\label{fig_completeness}
\end{figure*}

\begin{figure*}[ht!]
\sidecaption
\includegraphics[width=12cm]{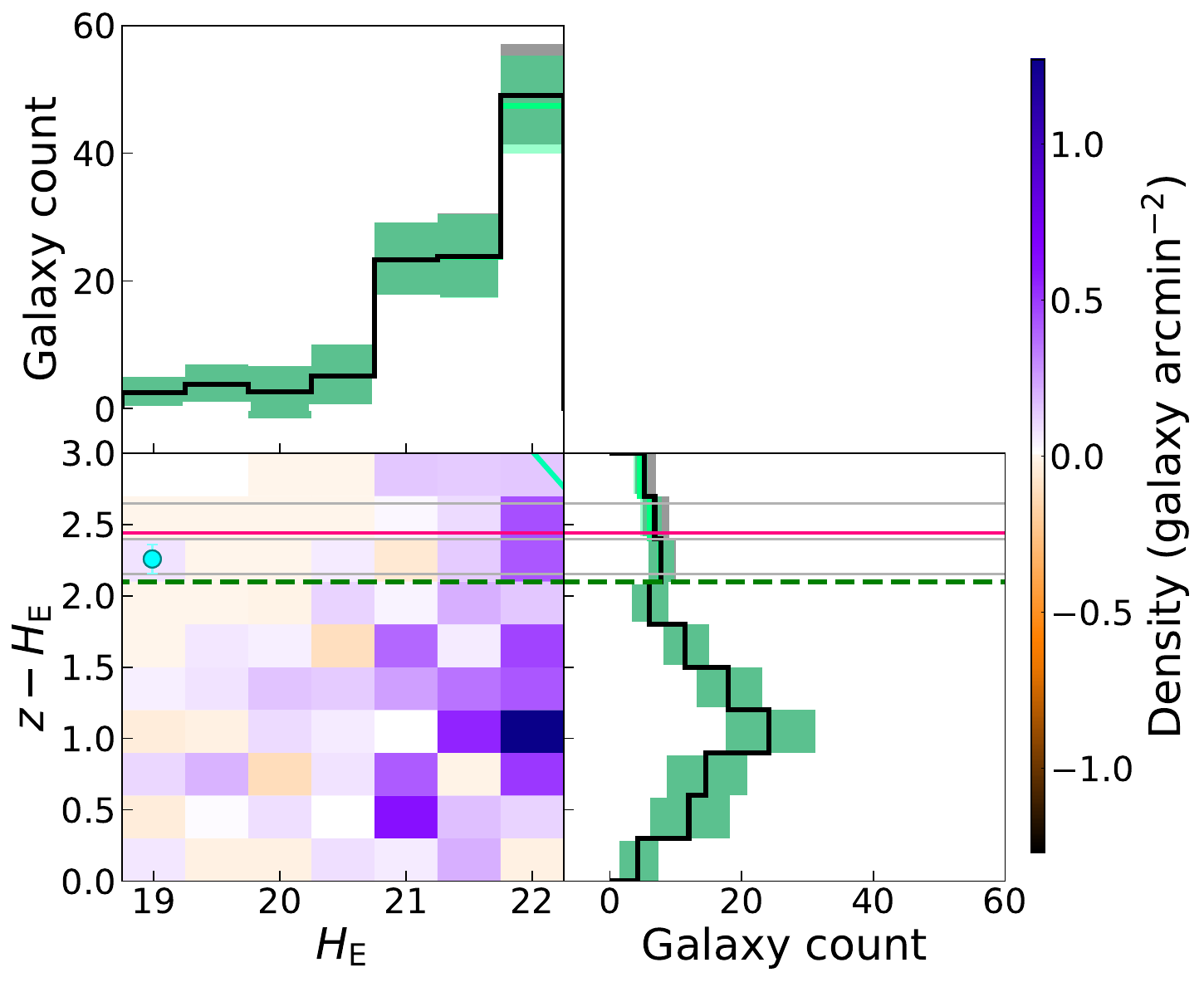}
\caption{\emph{Bottom left}: Field density-subtracted, completeness-corrected CMD. Each colour encodes a galaxy overdensity (purple) or underdensity (orange), expressed in count per arcminute square. We restricted the magnitude to $\HE \leq 22.25$. As in Fig. \ref{fig_cmd}, the locations of the red sequence at different redshifts are indicated by grey lines, and the location at $z = 1.74$ is indicated in pink. On the $z=1.74$ red sequence, $\HE= 22.25$ corresponds to $z=24.69$ and to a completeness level of 76\%. The cyan line on the very top right on the panel indicates the 50\% $z$ completeness limit and the dashed green line shows the location of the red sequence/blue cloud boundary. The proto-BCG photometry is marked by a cyan dot. \emph{Top Left}: Count of member galaxies as a function of \HE-band magnitude, after subtraction of the background level. The solid green and black step lines indicate the background-subtracted counts before and after the application of the completeness correction, while the shaded regions indicate count uncertainties. \emph{Bottom Right}: Background-subtracted members count as a function of colour with and without the completeness correction.}
\label{fig_cmd_subtracted}
\end{figure*}

Spectroscopic data were reduced with the \texttt{PypeIt} data reduction pipeline version 1.17.1 \citep{prochaska_pypeit_2020}, following the standard calibration and extraction process. The reduction process included bias subtraction, flat-field correction, and wavelength calibration employing arc lamp frames. Sky subtraction was performed using the \texttt{PypeIt} algorithm, which models the sky background on a slit-by-slit basis to account for variations across the field of view. \texttt{PypeIt} outputs reduced and co-added 2D spectra, from which we extract 1D spectra using a constant extraction width as a function of wavelength with an object-finding algorithm. The extraction width is calculated based on a \texttt{PypeIt}-based optimal aperture mask maximising the S\slash N. We examined the 1D and 2D spectra to locate emission lines.

\section{Analysis}\label{sec_analysis}

The detection of the Puddle cluster in MaDCoWS2 and \textit{Euclid} \citep{Q1-SP050} indicated the presence of a sizeable overdensity. MOSFIRE spectroscopy provided spectroscopic redshifts centred at $z\sim 1.74$ for the BCG (see Sect. \ref{ssec_spectro}) and for a galaxy at $z=1.769$ $\ang{;3.1}$ away from the BCG complex, which may be part of the cluster infalling region. In this section, we explored the candidate cluster properties with the photometric and spectroscopic data sets. When necessary, we assumed that the redshifts of the cluster and the BCG are $z = 1.74$.

\subsection{The galaxy population}\label{ssec_cmds}

Figure \ref{fig_cmd} presents the $z-\HE$ colour-magnitude diagram (CMD) of the Puddle cluster. The galaxies out to $\ang{;1}$ away (about $0.52\,$Mpc) from the proto-BCG are shown as green points, and those between $\ang{;1}$ and $\ang{;2}$ are in black. The red sequence colour is 2.44 at $z=1.74$ (pink line), assuming Solar metallicity. It is computed with \texttt{Bagpipes} \citep{carnall_inferring_2018} with a model consisting of a single burst of star formation at $z_\mathrm{form}=5$ followed by passive evolution. Combining this model with the \citet{mancone_formation_2010} luminosity function for the IRAC $3.6\,\micron$ channel, we compute a characteristic magnitude ($m^\ast$) of 21.4 in \HE band. A formation redshift of 3 generates similar results: a red sequence at $z-\HE=2.22$ and $m^\ast=21.2$.

The central structure is detected as four galaxies by the MER pipeline. However, given the irregular shape and the presence of substantial diffuse emission, the MER photometry is unreliable for these four sources. Thus, we removed those four detections from our CMDs and performed our own measurement of the proto-BCG flux in Sect. \ref{ssec_bcg}. The cyan dot in Fig. \ref{fig_cmd} represents the position of the central structure in the CMD.

The red sequence of our CMD appears sparsely populated compared with clusters confirmed at $z>1.5$ \citep[e.g.][]{papovich_spitzer-selected_2010,andreon_jkcs_2014,webb_extreme_2015,strazzullo_red_2016,willis_spectroscopic_2020}. We see in particular a lack of red galaxies at $\HE<21.4$, which is unusual among the known high-redshift clusters: massive galaxies are more susceptible to internal quenching \citep[e.g.][]{peng_mass_2010} and thus tend to be redder than less massive ones even at $z\sim 2$ \citep[e.g.][]{kawinwanichakij_effect_2017}.

\begin{table*}
\caption{Results of the decomposition of the main nucleus spectrum.}
\smallskip
\label{table_decomposition}
\smallskip
\centering
\begin{tabular}{cccccc}
\hline
& & & \\[-9pt]
Decomposition & Line & $z$ & FWHM & $\chi_\nu^2$ & AIC\\
&&&(km$\,\mathrm{s^{-1}}$)&&\\
\hline
One Gaussian & H$\alpha$ & $1.7408\pm 0.0007$ & $1669\pm 189$ & 1.82 & 3263\\
\hline
\multirow{2}{*}{Two Gaussians} & H$\alpha$ & $1.7357\pm 0.0005$ & $914\pm 151$ & \multirow{2}{*}{1.61} & \multirow{2}{*}{3190}\\
& H$\alpha$ & $1.7455\pm 0.0002$ & $431\pm 60$ &\\
\hline
\multirow{3}{*}{Three Gaussians} & [\ion{N}{ii}] & \multirow{3}{*}{$1.7368\pm 0.0001$} & $434\pm 56$ & \multirow{3}{*}{1.62} & \multirow{3}{*}{3194}\\
& H$\alpha$ & & $704\pm 146$ &\\
& [\ion{N}{ii}] & & $431\pm 56$ &\\
\hline
\end{tabular}
\end{table*}

\subsubsection{Background subtraction and completeness correction}\label{sssec_bck}

To assess the importance of the galaxy overdensity associated with the Puddle cluster and its location in the CMD diagram, we estimated the contribution of field galaxies and subtracted it from our CMD. To measure this `background', we randomly selected 100 non-overlapping circular regions ($r=\ang{;2}$) in the Fornax field. We computed the average density of galaxies in these fields as a function of \HE magnitude and $z-\HE$ colour and we used only the galaxies that are at least $3\,\sigma$-detected in both $z$ and \HE bands. For each \HE magnitude and colour, we then subtracted this average density from the galaxy density in the central region of the Puddle cluster. The background subtraction uncertainties were determined from the standard deviation of the galaxy densities in the random fields. 

We also used the 100 random fields described above to determine completeness corrections. Our computation followed the same principle than the completeness correction described in the Appendix A.1. of \citet{van_der_burg_gogreen_2020}. Figure \ref{fig_completeness} presents a visual summary of the process. We fitted an exponential to the highly complete part of the distribution in each band empirically determined as $z<23.5$ and $\HE<23$. It should be noted that since the $z$-band completeness is the most critical one, we have been especially conservative in our selection of the regime which we consider highly complete. We then computed a completeness factor for magnitudes greater than 23 (\HE) or 23.5 ($z$) by taking the ratio of the observed galaxy count to the count predicted by the best fit. We assumed a completeness factor of one for $z\leq 23.5$ and $\HE \leq 23$. The total 50\% completeness, shown on Fig. \ref{fig_cmd} as a purple line, is estimated by multiplying the $z$ and \HE-band completeness factor for a given set of $z$ and \HE magnitudes. This multiplication relies on the assumption that there is no correlation between detections in the $z$ and \HE bands. Since that may not be true, we limited the rest of our analysis to $\HE<22.25$, a regime where the completeness correction does not fall below 34\% and is entirely determined by the $z$-band completeness (i.e. the completeness correction is entirely determined by colour). The completeness uncertainties were propagated from the uncertainties on the best fit parameters. For example, the correction for a $z$-band completeness of 0.5 is $2.0\pm 0.2$.

Figure \ref{fig_cmd_subtracted} presents the impact of these corrections on the galaxy population of the Puddle cluster candidate. The bottom left panel shows the background-subtracted, completeness-corrected distribution. The 50\% $z$-band completeness limit is shown as a cyan line. To avoid large (and uncertain) completeness corrections, we limit the CMD in the bottom left panel to $\HE\leq 22.25$ ($\HE=22.25$ is equivalent to $z=24.69$ on the red sequence and to a 76\% complete $z$ band). We limited the colour to $z-\HE=3$, as the background-subtracted galaxy count is consistent with zero for redder colours. The other panels show the galaxy count before and after the application of the completeness correction as a function of \HE (top left) and $z-\HE$ colour (bottom right). For $\HE\leq 22.25$, the excess count within $\ang{;2}$ of the BCG is $108\pm 14$ galaxies above the field level. It increases to $110\pm 14$ galaxies after the application of the completeness correction. The red sequence location is indicated on the top left and bottom panels. The count uncertainties are based on the propagation of the density subtraction and the completeness uncertainties when relevant.

As noted above, there are $18\pm 3$ galaxies ($20\pm 4$ after the completeness correction) in excess compared to the field level in the red sequence and very few luminous red galaxies outside of the BCG complex ($1\pm 1$ galaxies with $\HE<21$). The bulk of the overdensity is about 1.5 magnitudes bluer than the red sequence population, and fainter than $\HE=20$. With \texttt{Bagpipes}, we computed that $z-\HE=0.94$ corresponds to a $\sim 155\,$Myr old stellar population, assuming Solar metallicity and no attenuation. While this blue colour suggests widespread star formation, we did not attempt to determine the cluster specific SFR. The lack of bright red galaxies and the imprecise photometric redshifts of \Euclid at $z\gtrsim 1.5$, are probably the two main factors explaining the difference between our galaxy counts and the \citet{Q1-SP050} richness estimate.

\subsection{Spectroscopy}\label{ssec_spectro}

\begin{figure}[ht!]
\vspace{-20pt}
\centering
\includegraphics[angle=0,width=\columnwidth]{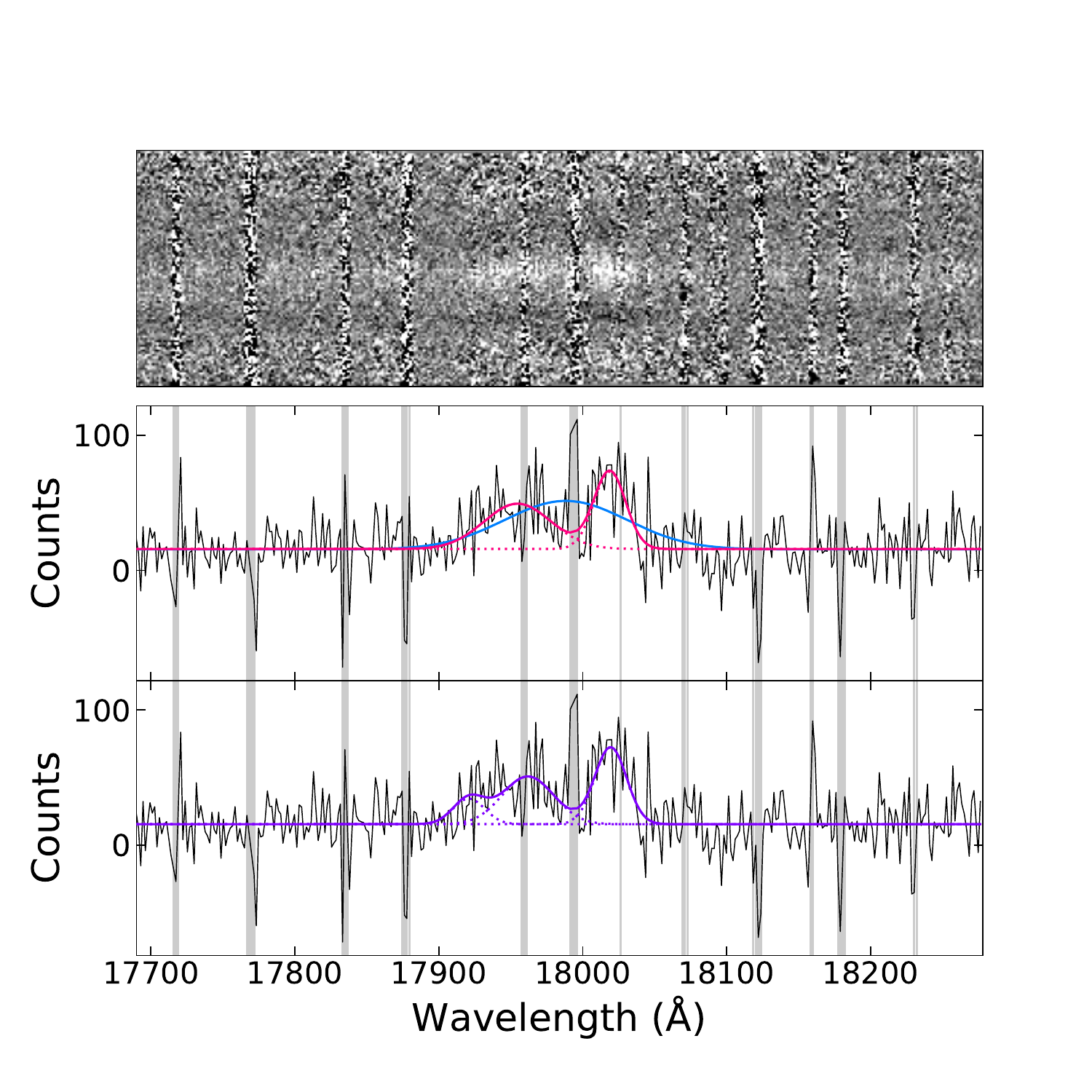}
\caption{\emph{Top}: Portion of the proto-BCG 2D spectrum from MOSFIRE, showing emission lines. The spectrum shows the entire slit height but only a part of the spectral range, centred on the emission lines. \emph{Middle}: Same portion of the 1D spectrum. The blue and pink lines show possible decompositions into one and two H$\alpha$ emission lines respectively. \emph{Bottom}: Same as in the middle panel, but presenting a third possible decomposition (in purple): H$\alpha$ with the [\ion{N}{ii}] doublet on each side. The dotted lines trace the individual components of each fit. The shaded zones on the 1D spectra correspond to the regions with the 10\% lowest weights in the fits, i.e. regions with subtracted sky lines. The fits are performed on the 1D spectrum.}
\label{fig_decomposition}
\end{figure}

We detected continuum emission of varying robustness in 14 of the 18 MOSFIRE slitlets. We also identified unambiguous isolated emission lines, most likely H\,$\alpha$, in three slitlets: one slitlet targeting the BCG complex, one slitlet targeting a source approximately $\ang{;3.5}$ from the MaDCoWS2 centroid, and another slitlet far from the MaDCoWS2 centroid that has two serendipitous emission lines. In addition, faint, tentative emission lines were detected in three other slitlets. Based on one of these faint tentative line, we estimated a line detection threshold of $\sim 4\, \times\, 10^{-17}\, {\rm erg}\, {\rm cm}^{-2}\, {\rm s}^{-1}$ at the central wavelengths of the $H$-band grism. This corresponds to a star-formation rate of $\sim 5\, M_\odot\, \mathrm{yr^{-1}}$, assuming a standard conversion factor between star-formation rate and H\,$\alpha$ line luminosity \citep{hao_dust-corrected_2011}. At $\sim 1.8\, \mu$m, the wavelength of the emission line detected from the BCG complex, the atmosphere absorbs $\sim 60\%$ of the flux, implying a rough star-formation rate detection threshold of 8$\,M_\odot\,\mathrm{yr^{-1}}$. The presence of faint traces and tentative emission lines suggests that the exposure time calculations were perhaps too optimistic. We also note that the outcome of our observation is comparable to \citet{stanford_idcs_2012} attempt to confirm IDCS J1426.5+3508, a $z=1.75$ cluster, with spectra from the Low-Resolution Imaging Spectrograph on \textit{Keck I}. On 16 targets, they confirmed two members. However, the main challenge in our data reduction was the subtraction of the strong near-infrared sky lines \citep[e.g.][]{webb_extreme_2015,balogh_gemini_2017,belli_mosfire_2019}, while \citet{stanford_idcs_2012} main limitation was the faintness of quiescent galaxies in the optical (which corresponds to rest-frame ultraviolet at $z\sim 1.75$).

Figure \ref{fig_decomposition} presents a portion of the spectrum of BCG complex, centred on the emission line. As shown by the top panel, the emission originates solely from the proto-BCG central nucleus. While the slit was designed to be centred on the main core and to include the southwestern nucleus (see the right panel of Fig. \ref{fig_presentation}) we do not see any clear secondary trace. It is possible that the southwestern nucleus may possess only a very faint continuum and no significant emission lines; we note also that the extensive diffuse emission in the BCG complex may have been subtracted as sky emission, resulting in a lower signal in the vicinity of the southwestern nucleus. Alternatively, if a small misalignment occurred during observations, the second nucleus may have fallen outside of the slit during some of the dithers. The spectral resolution varies across the spectrum, but is on average $1.6\,$\AA$\,$pixel$^\mathrm{-1}$.

The emission region in Fig. \ref{fig_decomposition} is broad, spanning roughly $150\,$\AA~and presents a complex structure. Such a broad range suggests three possibilities: a single H$\alpha$ line from an AGN; at least two sources of H$\alpha$ emission along the same line-of-sight; or H$\alpha$ emission blended with the [\ion{N}{ii}] doublet. We examined these possibilities by modelling the spectrum as one, two, and three Gaussians, weighting the contribution of each pixel by the inverse of its variance. We computed the reduced $\chi^2$ and Akaike information criterion \citep[AIC, e.g.][]{akaike_new_1974,liddle_information_2007} of each fit. The results are presented in Table \ref{table_decomposition}. The middle and lowest panels of Fig. \ref{fig_decomposition} show the one-, two-, and three-Gaussian decompositions in blue, pink, and purple, respectively. Due to the limited S\slash N of the spectrum, we did not attempt to subdivide the H$\alpha$ emission into broad and narrow components.

The simplest case, the single Gaussian model, yields the highest reduced $\chi^2$ and AIC. Both measures strongly suggest that the two- or three-Gaussians models are better representations of the spectrum. The components of the three-Gaussian model are not treated independently: We fixed the amplitude ratio of the [\ion{N}{ii}] doublet as $1/3$, in agreement with theoretical \citep[e.g.][]{galavis_atomic_1997,storey_theoretical_2000,bon_ratios_2025} and observational \citep{dojcinovic_flux_2023} determinations, and forced its members to have the same width. We forced all three emission lines to have the same redshifts. We also tested a fit in which we relaxed the last requirement and allowed the H$\alpha$ emission to lie at a different redshift than the [\ion{N}{ii}] doublet. We found very similar redshifts ($\Delta z=0.0002$) and the FWHM of each line was not significantly changed compared to the previous fit. We discussed the interpretation of the various fits in Sect. \ref{sssec_sf}.

\begin{figure*}[hbt!]
\centering
\includegraphics[angle=0,width=\textwidth]{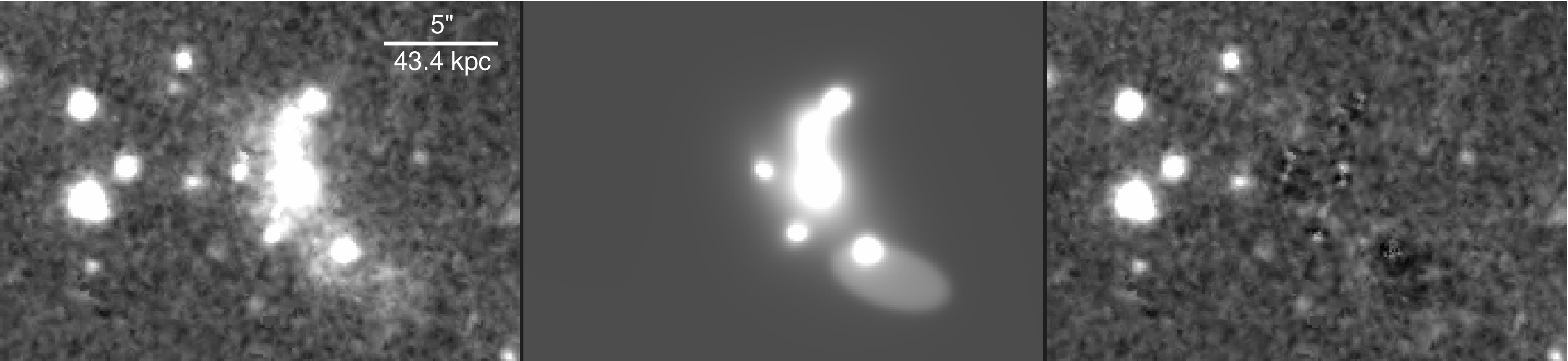}
\caption{\emph{Left}: \HE-band image of the central merger. \emph{Middle}: Central merger model, made of nine S\'ersic profiles. \emph{Right}: Residuals. The luminosity scale is the same for the three panels. North is up, and east is left.}
\label{fig_puddle_galfit}
\end{figure*}

\subsection{The BCG complex}\label{ssec_bcg}

The right panel of Fig. \ref{fig_presentation} presents a tricolour image (\IE, \YE, and \HE) of the BCG complex of the Puddle cluster, with the MOSFIRE slit and the custom aperture used for the photometry overlaid (see Sect. \ref{sssec_phot}). This system has a complex morphology, with diffuse emission embedding at least six cores and extending in a kind of tail towards the southwest. There is a seventh core within the tail emission. The projected distance between cores is $\lesssim 72\,$kpc, and in the case of the six central cores $\lesssim 65\,$kpc. Along the axis of the diffuse tail, the system is $\sim 105\,$kpc across.

\subsubsection{\textit{Euclid} and DECam photometry}\label{sssec_phot}

To estimate the magnitude of the BCG complex, our first step was to determine a suitable aperture to measure the system photometry. To do so, we used \texttt{sep.extract} to compute the segmentation map in the NISP \HE band and set our detection threshold at $3\,\sigma$ with a minimum area of 30 pixels. We then assigned a value of one to the pixels corresponding to the BCG complex detections and removed all other detections in our segmentation map. The resulting detection was very irregularly shaped, so we used a Gaussian kernel with $\sigma=$ 3 pixels to smooth the detection shape, retaining only the pixels with a value of 0.25 or more. We then used the resulting shape (shown as white contours in the right panel of Fig. \ref{fig_presentation}) to measure the photometry in all \textit{Euclid} and DECam bands.

To estimate the photometric uncertainties, we adopted a statistical approach: We found all detections in the image of interest, with a detection threshold at $3\,\sigma$ and a minimum area of 5 pixels. We then randomly displaced the footprint of the central merger until we found a position with no overlapping detections. We measured the sky flux in that position and repeated the process for 100 random empty sky patches. We took the standard deviation of the sky fluxes as our sky uncertainties, which we then added in quadrature to the Poisson noise of the aperture photometry.

We did not compute aperture corrections for DECam. As long as the defined aperture contains most of the flux in every band, the aperture corrections should be negligible. This is the case for all DECam bands. In 3$\,\sigma$ segmentation maps, the BCG complex, if detected, is entirely enclosed by our custom aperture.

\subsubsection{IRAC photometry and morphology modelling}\label{ssec_galfit}

The pixel size of the IRAC mosaics is $\ang{;;0.6}$, which is six times larger than the the pixel size of \textit{Euclid} VIS and NISP images. To measure the flux of the marginal detections in the IRAC $5.8$ and $8.0\,\micron$ channels, we resampled our convolved \HE segmentation map to the correct pixel size (setting the pixels with fractional values to unity) and performed the flux and uncertainty measurements as described above. 

However, at $3.6$ and $4.5\,\micron$ the BCG complex is heavily blended with neighbouring galaxies. We used \texttt{Galfit} modelling \citep{peng_detailed_2002,peng_detailed_2010} to disentangle the structure from neighbouring galaxy fluxes. 
Using the high resolution of the NISP data \citep{fazio_infrared_2004,EuclidSkyNISP}, we constructed initial models in \HE band. We first modelled the BCG complex with seven S\'ersic profiles for the structure nuclei, and two additional S\'ersic profiles for the diffuse emission. Because no PSF estimation is provided with Q1 frames, we approximated \textit{Euclid} PSFs using a neighbouring unsaturated star. The result is shown in Fig. \ref{fig_puddle_galfit}. We then made a separate model of the contaminants before merging the two models together.

Our IRAC $3.6$ and $4.5\,\micron$ models are directly based on the combined \HE-band models, with a few changes. Unlike the \textit{Euclid} data, IRAC mosaics have uncertainty maps, which we added into the \texttt{Galfit} settings. We also deleted one of the two diffuse S\'ersic profiles, as we could not make the merged models converge otherwise. Likewise, we removed some of the faintest interlopers, as they were too faint to be detected in the IRAC images. For our first pass models, we only let the magnitudes change, but for our final models we allowed positional changes as well to account for shifts induced by PSF miscentring; the effective radii, S\'ersic indexes, axis ratios, and position angles were fixed to initial values.

We tested two different PSFs: an empirical PSF based on the same star used to estimate the \HE-band PSF, and a stacked theoretical point response function (PRF). The stacked PRF was made by rotating the model PRF of the cryogenic mission\footnote{see \url{https://irsa.ipac.caltech.edu/data/SPITZER/docs/irac/calibrationfiles/psfprf/}} to the position angles of all the input images contributing to the mosaic at the location of the BCG. The rotated PRFs were weighted by the square root of the exposure times and stacked. The empirical PSF performed better than the stacked PRF, which was too concentrated.

We measured the flux of the BCG complex in the IRAC $3.6$ and $4.5\,\micron$ channels directly from the best \texttt{Galfit} models. To ensure consistency with the aperture photometry (see Sect. \ref{sssec_phot}), we computed an aperture correction by dividing the aperture flux of the \HE band by the \texttt{Galfit} flux in the same band. We found an aperture correction of $0.83 \pm 0.05$.

\subsubsection{Spectral energy distribution fitting}\label{sssec_SED}

\begin{table}[htbp!]
\caption{Photometry used for SED modelling.}
\smallskip
\label{table_photometry}
\smallskip
\centering
\begin{tabular}{ccc}
\hline
& \\[-9pt]
Instrument & Band & Flux\\
 & & ($\mathrm{\mu}$Jy)\\
 & \\[-9pt]
\hline
DECam & $g$ & $2.5\pm 0.4$\\
DECam & $r$ & $4.1\pm 0.4$\\
DECam & $i$ & $6.9\pm 0.7$\\
DECam & $z$ & $12\pm 1$\\ 
VIS & \IE & $6\pm 2$\\
NISP & \YE & $39\pm 2$\\
NISP & \JE & $62\pm 1$\\
NISP & \HE & $92\pm 1$\\
IRAC & $3.6\,\micron^1$ & $196\pm 25$\\
IRAC & $4.5\,\micron^1$ & $228\pm 20$\\
IRAC & $5.8\,\micron$ & $155\pm 24$\\
IRAC & $8.0\,\micron$ & $90\pm 18$\\
MIPS & $24\,\mathrm{\micron}$ & $205\pm 59$\\
\hline
\end{tabular}
\justify

$^1$ After multiplication by an aperture correction factor of $0.83\pm 0.05$. See the main text for more details.\\
\end{table}

Table \ref{table_photometry} presents the derived DECam, \textit{Euclid}, and IRAC photometry, which we used to perform spectral energy distribution (SED) modelling of the entire BCG complex. We did not attempt to model the SED of individual cores: The signal-to-noise of the \IE and DECam fluxes was not sufficient to perform a \texttt{Galfit} decomposition.

\begin{figure}
\centering
\includegraphics[angle=0,width=0.99\columnwidth]{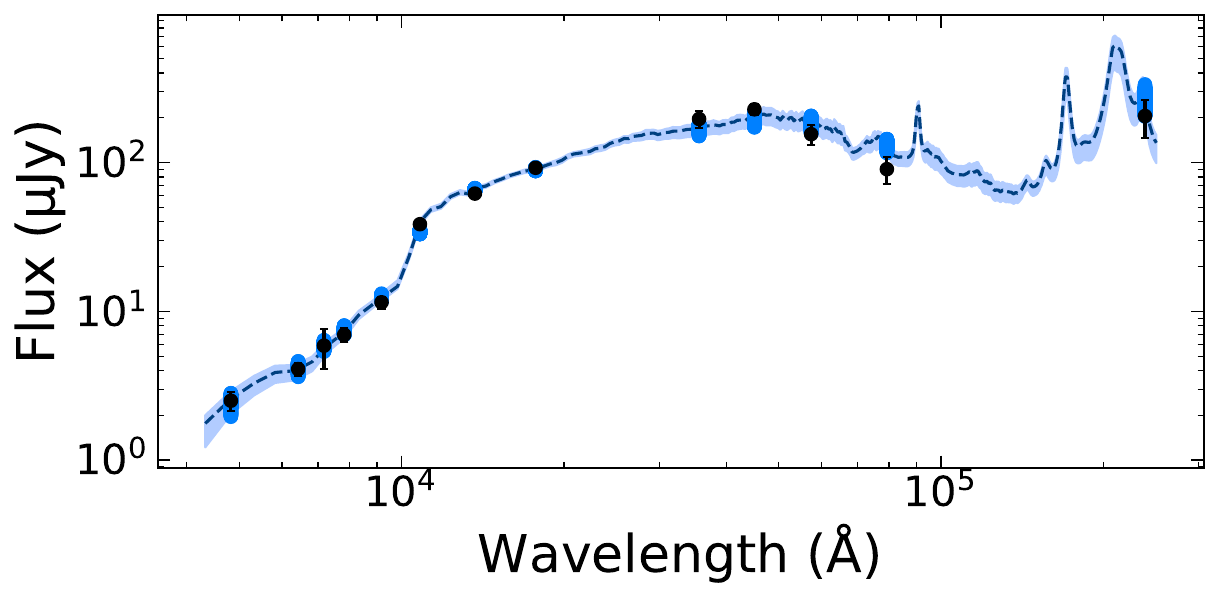}
\caption{Results of the SED modelling of the central structure. The shaded area corresponds to the SEDs within the 68\% confidence interval, and the blue circles to the expected photometry for each of these models. The black dots are the measured fluxes. The most likely model (dark dashed line) has a reduced $\chi^2$ of 3.34.}
\label{fig_puddle_sed}
\end{figure}

We adopted most of the \texttt{Bagpipes} \citep[][]{carnall_inferring_2018} standard basic assumptions \citep[models from][with a \citealt{kroupa_mass_2002} initial mass function, uniform priors]{bruzual_stellar_2003}, and added a dust component that follows the \citet{calzetti_dust_2000} attenuation law. We replaced the \texttt{Bagpipes} default cosmology with the \citet{collaboration_planck_2020} cosmological parameters. We adopted a parametric approach, testing a delayed $\mathrm{\tau}$-model and a model in which all the stars formed at the same time (i.e. a burst). The burst yields a best fit SED with a smaller reduced $\chi^2$ than the delayed $\mathrm{\tau}$-model (3.34 compared to 3.75). Furthermore the characteristic time of the delayed $\mathrm{\tau}$-model is extremely short, $60^{+50}_{-40}\,$Myr. The free parameters, priors, and results of the bursty model are presented in Table \ref{table_SED}. The results are expressed as the median of the parameter probability distribution, and their uncertainties were estimated using the 16 and 84th percentiles. The resulting SED is shown in Fig. \ref{fig_puddle_sed}. Note that \texttt{Bagpipes} fits the total stellar mass formed. The current stellar mass is $(5.7\pm 0.3)\times 10^{11}\,M_\odot$, and is also included in Table \ref{table_SED}.

We also tested more complicated models: first two bursts and then a burst with a delayed $\mathrm{\tau}$-model, assuming Solar metallicities to reduce the number of free parameters. In both cases, the results were very poorly constrained.

In all these fits, we assumed that the central AGN contribution to the flux is minimal. To assess the impact of the AGN emission on the fit, we used the AGN module in \texttt{Bagpipes} \citep{carnall_massive_2023}, which implements a  \citet{vanden_berk_composite_2001} quasar model. The model consists of two power laws describing the continuum blueward and redward of 5100$\,$\AA~in the rest frame, normalised by the flux of the AGN at 5100$\,$\AA~and by the flux and width of the H$\alpha$ emission line. We tested different variations of this model, assuming that the AGN is responsible for 10\% of the flux at 5100$\,$\AA~(which we estimate using the most likely SED of the fit without an AGN) and that the flux of the H$\alpha$ emission line corresponds to 10\% of the \HE band flux. We obtained large reduced $\chi^2$ values. We then let the 5100$\,$\AA~normalisation vary and found that the most likely AGN contribution to the flux at 5100$\,$\AA~is negligible ($0.09^{+0.12}_{-0.07}$\%) and that the other parameters of the fit corresponded to the values listed in Table \ref{table_SED}. We thus concluded that the AGN does not contribute significantly to the SED of the BCG complex.
 
\begin{table}
\caption{Fitting parameters, priors, and results of the SED fit.}
\smallskip
\label{table_SED}
\smallskip
\begin{tabular}{ccc}
\hline
& \\[-9pt]
Parameter & Prior$^{1}$ & Results\\
& & \\[-9pt]
\hline
& & \\[-9pt]
Age (Gyr) & [0.001,3.74]$^{2}$ & $0.32^{+0.04}_{-0.03}$\\
& & \\[-9pt]
Mass formed [$\logten(M/M_\odot)$] & [10,13] & $11.94^{+0.03}_{-0.02}$\\
& & \\[-9pt]
A$\mathrm{_v}$ (mag) & [0,2] & $0.95^{+0.08}_{-0.10}$\\
& & \\[-9pt]
Metallicity ($\mathrm{Z/Z_\odot})$ & [0,2.5]$^{3}$ & $1.11^{+0.48}_{-0.40}$\\
& & \\[-9pt]
\hline
& & \\[-9pt]
Stellar mass [$\logten(M/M_\odot)$]$^4$ & - & $11.76^{+0.03}_{-0.02}$\\
& & \\[-9pt]
\hline
\end{tabular}
$^1$ All priors are uniform.\\
$^2$ The code automatically limited the prior to the age of the Universe at the considered redshift.\\
$^3$ We also tested models with fixed metallicities. See the explanation in the main text.\\
$^4$ \texttt{Bagpipes} does not fit directly the current stellar mass, but rather the total stellar mass formed during the star-formation history. We included the current stellar mass for indicative purposes only.\\
\end{table}

\section{Discussion}\label{sec_discussion}

\subsection{Cluster red fraction}\label{ssec_cluster_discussion}

The Puddle is a candidate cluster detected in two different surveys, MaDCoWS2 and \textit{Euclid} \citep{thongkham_massive_2024b,Q1-SP050}. It still corresponds to a significant galaxy overdensity after subtraction of the mean galaxy density in the Euclid Deep Fornax Field (see Sect. \ref{ssec_cmds}). While a single redshift is not enough to spectroscopically confirm a cluster \citep[e.g.][]{eisenhardt_clusters_2008,noirot_hst_2018}, the facts enumerated above suggest that the Puddle cluster is a genuine high-redshift cluster (see also the characterisation of similar unconfirmed clusters by \citealt{valtchanov_xmm-lss_2004} and \citealt{cooke_mature_2016}).

As noted in Sect. \ref{ssec_cmds}, the Puddle cluster red sequence is relatively sparse compared to other $z>1.5$ clusters and dominated by faint red members. This situation would make the detection of the Puddle cluster challenging with the red-sequence method \citep[e.g.][]{gladders_new_2000,gladders_red-sequence_2005,donahue_distant_2002}, one of the commonly used algorithms for optical/near-infrared cluster searches \citep[e.g. the C4 clustering algorithm, the Spitzer Adaptation of the Red-sequence Survey and the red-sequence Matched-filter Probabilistic Percolation;][]{miller_c4_2005,wilson_clusters_2006,wilson_spectroscopic_2009,rykoff_redmapper_2014}.

While Fig. \ref{fig_cmd} does not show a clear division between the red sequence and the blue cloud, in the colour histograms shown in Fig. \ref{fig_cmd_subtracted} bottom right panel, there is a local minimum at $1.8< z-\HE\leq 2.1$. We thus adopted $z-\HE=2.1$, which corresponds to a stellar age of 1.08$\,$Gyr (Solar metallicity, no attenuation), as our limit between the red sequence and the blue cloud. With this assumption, we found, within a \ang{;2} radius centred on the BCG complex, $17\pm 3$ red galaxies ($\HE<22.25$) without completeness correction, and $18\pm 4$ red galaxies with it. In both cases, the count of blue galaxies was $90\pm 14$. These counts correspond to red fractions of $17\pm 4$\% and $18\pm 4$\% respectively. The uncertainties were based on the propagation of the uncertainties of the background subtraction and the completeness correction (see Sect. \ref{sssec_bck}).

This would make the red fraction of the Puddle cluster consistent with the measurements of \citet{quadri_tracing_2012} and \citet{nantais_stellar_2016} in $z\sim 1.6$ clusters, but significantly less quenched than the clusters of \citet{newman_spectroscopic_2014}, \citet{cooke_mature_2016}, and \citet{strazzullo_galaxy_2019}. However, the comparison with other works is made challenging by the different stellar mass, magnitude, and radius cuts applied by each author, and by the dust obscuration, which can be mistaken for quiescence \citep[e.g.][]{woo_dependence_2013,hatch_structure_2016}.

In particular, our own magnitude cut does not correspond to a uniform stellar mass cut. At similar \HE-band magnitudes, blue galaxies will have a lower stellar mass than red members. Assuming a simple stellar population model with no reddening, we computed approximate stellar mass as a function of colour and magnitude and found no overlap between the mass range probed by the reddest and the bluest colour in our CMD. To enable a uniform stellar mass cut, we would need deeper $z$-band data than those currently available.

\subsection{BCG formation scenarios}\label{ssec_scenarios}

The morphology of the BCG complex is consistent with a complex multiobject merger. While line-of sight superpositions remain possible, the six central cores are embedded in diffuse emission, which suggests that they are interacting. The southwestern core may or may not be part of the merger. The extension of the diffuse emission to the southwest is reminiscent of a tidal tail \citep[see e.g.][]{webb_extreme_2015}. We thus inferred that the Puddle cluster is in the process of forming its BCG through a multiobject merger. In this section, we explored the properties of this merger and how it compares with SPT2349$-$56 \citep{miller_massive_2018} and SpARCS104922.6+564032.5 \citep{webb_extreme_2015}. We noted two other instances of multiobject mergers interpreted as assembling BCGs in the literature: the Spiderweb protocluster at $z=2.16$ \citep{miley_spiderweb_2006} and an unnamed $z=1.85$ overdensity in the Cosmic Evolution Early Release Science of JWST \citep{coogan_z_2023}.

\subsubsection{Star formation and AGN activity}\label{sssec_sf}

Most of the proto-BCGs in the literature exhibit high levels of star formation, obscured or not \citep{seymour_rapid_2012,webb_extreme_2015,miller_massive_2018,coogan_z_2023}. In the Puddle cluster, the SED model that best fits the photometry of the BCG complex is an instantaneous burst of star formation $320\,$Myr ago, with no more recent star formation. We note however that parametric SED modelling should be taken with caution \citep[e.g.][]{carnall_how_2019,leja_how_2019,haskell_beware_2024,mosleh_reconstructing_2025}.
Thus, we interpret these results as evidence for a burst of star formation about $300\,$Myr ago, but do not rule out the possibility of more recent, residual star formation. In fact, given the young age of the dominant stellar population, ongoing unobscured star formation in the Puddle cluster proto-BCG is plausible. Obscured star formation is also possible: All star-formation histories returned large dust attenuations, regardless of the metallicity content. Without coverage in the far infrared, it is difficult to assess its importance.

The spectrum of the brightest nucleus does not offer any clues about the current SFR since it is dominated by AGN emission. If we model the spectrum as two Gaussians (see Sect. \ref{ssec_spectro}), the broadest emission line FWHM is $914\,\mathrm{km\,s^{-1}}$ (i.e. $\sigma=388\,\mathrm{km\,s^{-1}}$). This is too large to be associated with the merging activity, since the line width is similar to the total velocity dispersion of an entire (proto)cluster \citep[e.g.][]{kuiper_sinfoni_2011,ruel_optical_2014}.

The results of the three-Gaussian modelling also suggest that the spectrum is AGN-dominated. Using this model, we computed approximate equivalent-width ratios of $\logten(\mathrm{([N\,II]\lambda 6548+[N\,II]\lambda 6584)}/\mathrm{H\,}\alpha)=0.12$ and $\logten(\mathrm{[N\,II]\lambda 6584}/\mathrm{H\,}\alpha)=0.00$. Large values like these are unlikely to be associated with star formation alone, but consistent with the ionisation provided by an AGN \citep[e.g.][]{bian_direct_2018,oh_observational_2019,oh_bass_2022,garg_bpt_2022,dors_cosmic_2023,EP-Scharre,zhou_mammoth-mosfire_2025}.

The stellar mass computed by Bagpipes is $(5.7\pm 0.3)\times 10^{11}\,M_\odot$. This value does not include AGN contamination and assumes a single stellar population, about 300$\,$Myr old. An AGN can sometimes result in an overestimation of the stellar mass \citep[e.g.][]{ciesla_constraining_2015,salim_galex-sdss-wise_2016}. However, the presence of an older population would instead decrease the mass-to-light ratio \citep[e.g.][]{mancone_ezgal_2012,conroy_modeling_2013}. Bagpipes stellar mass should thus be treated as an estimate.

\subsubsection{Comparison with SPT2349$-$56 simulations}\label{sssec_SPT2349}

Lying at $z=4.3$, SPT2349$-$56 is a protocluster with 14 massive gas-rich galaxies all within $130\,$kpc of projected distance from each other \citep{miller_massive_2018}. According to the simulations of \citet[][see also \citealt{sulzenauer_bright_2026} updated versions]{rennehan_rapid_2020}, SPT2349$-$56's 14 central galaxies should merge into a single BCG in $\sim\,370\,$Myr from the time of observation. They found that the star-formation rate peaks very early in their simulation, before decreasing exponentially with a characteristic time of $200\,$Myr. \citet{remus_young_2023} comparison with the Magneticum cosmological simulations also predicts a rapid collapse.

The proto-BCG of the Puddle cluster has not yet coalesced into a single galaxy. However, the age and burst-like star-formation history of the unobscured stellar population suggest that we could be witnessing a more advanced stage of an SPT2349-like merger. The more compact morphology of the Puddle cluster's proto-BCG (see Sect. \ref{ssec_bcg}) also supports this hypothesis. Likewise, the stellar mass computed by \texttt{Bagpipes}, $(5.7\pm 0.3)\times 10^{11}\,M_\odot$, is larger than the initial stellar mass of the \citet{rennehan_rapid_2020} simulations ($2.81\times 10^{11}\,M_\odot$) and smaller than the predicted mass in $370\,$Myr ($7.33\times 10^{11}\,M_\odot$).

\citet{rennehan_rapid_2020} also made predictions about the probability of observing massive collapse events like SPT2349$-$56 (or the Puddle cluster) at different redshifts. While they observed events at all redshifts, most of them occured before $z\sim 2$. Yet, observationally, SPT2349$-$56 is the highest-redshift merging proto-BCG in the literature \citep{miller_massive_2018}. All the others lie at redshifts closer to the Puddle cluster -- between $z=1.71$ and $z=2.16$ \citep{miley_spiderweb_2006,webb_extreme_2015,coogan_z_2023}. A sample of five proto-BCGs is not enough to draw statistically significant conclusions. However, the fact that most of them lie at $z\lesssim 2$ suggests either that most collapses occur later and/or on longer timescales than the \citet{rennehan_rapid_2020} predictions, or that selection effects decrease the likelihood of detecting merging BCGs at $z\gtrsim 2$.

\subsubsection{Comparison with SpARCS104922.6+564032.5}\label{sssec_SpARCS1049}

While SPT2349$-$56 is the only proto-BCG with dedicated simulations, the SpARCS104922.6+564032.5 system is more directly comparable to the Puddle cluster in terms of redshift and physical scale. At $z=1.71$, SpARCS104922.6+564032.5's star-forming core is dominated by a $\sim 66\,$kpc-long tidal tail, with small clumps of various colours \citep{webb_extreme_2015}. It is unclear if the tidal tail is directly associated with the BCG, or with the pair of smaller merging cluster members located to its north. Subsequent studies have shown that the BCG hosts a faint radio-loud AGN \citep{trudeau_multiwavelength_2019} and that the bulk of the star formation occurs at $\sim12\,$kpc to the southeast of the BCG \citep{barfety_assessment_2022}. This corresponds to the centre of the tidal tail, where the MIPS emission of the system is located. The discovery of X-ray emission centred on the star-forming region suggests that this system might be powered by a displaced cooling flow rather than a merger \citep{hlavacek-larrondo_evidence_2020}.

The presence of a cooling flow in the Puddle cluster cannot be determined without X-ray or SZ observations. However, the Puddle cluster displays none of the clues that suggested the presence of a cooling flow as in SpARCS104922.6+564032.5. Unlike that cluster, the BCG of the Puddle cluster is clearly merging. Furthermore, the centroid of the MIPS emission is almost directly located onto the brightest of the merging galaxies. The proto-BCG emission lines are most likely powered by an AGN (see Sect. \ref{sssec_sf}) and are inconsistent with star formation as the primary ionising mechanism. The H$\alpha$ line is also too broad to be produced by a filamentary nebula similar to those associated with local cooling flows \citep[e.g.][]{mcdonald_origin_2010,hamer_optical_2016,gaspari_shaken_2018,gendron-marsolais_revealing_2018,olivares_ubiquitous_2019,olivares_gas_2022,mohapatra_velocity_2022}. Only the extreme cooling flow in the Phoenix cluster -- which is associated with a large amount of unobscured star formation \citep{mcdonald_massive_2012,mcdonald_hst/wfc3-uvis_2013} -- produces emission lines with FWHM around $1000~\mathrm{km~s^{-1}}$ \citep{reefe_directly_2025}. 

\subsection{Finding other assembling BCGs}\label{ssec_finding_others}

The discovery of an assembling proto-BCG in one of the two Euclid Deep Fields covered by MaDCoWS2 suggests that these objects might be common -- a hypothesis that is further supported by the presence of similar merging BCGs in the literature. Yet it is difficult to assess the prevalence of these systems because high-resolution near-infrared \citep[e.g. this study; see also][]{kuiper_sinfoni_2011,coogan_z_2023} or (sub)millimetre \citep[][see also \citealt{rotermund_optical_2021}]{miller_massive_2018} observations are necessary to reveal the complex structures of proto-BCGs. 

The large areas of the sky that will be observed by \textit{Euclid}, \textit{Nancy Grace Roman}, and \textit{Vera C. Rubin} should allow more systematic detections of merging proto-BCGs. In particular, the Euclid Wide Survey will cover approximately $14\,000\,$deg$^{2}$ at the same depth as the Q1 data release \citep{Scaramella-EP1,EuclidSkyOverview,Q1-TP001}. Thus, assuming that our single detection in the overlap between Euclid Q1 Data Release and MaDCoWS2 ($33\,$deg$^{2}$) is representative of the density of proto-BCGs on the sky, we would expect to find about 400 proto-BCGs in the completed Euclid Wide Survey.
 
\section{Conclusions}\label{sec_conclusion}

This work presents the characterisation of MOO2$\,$J03374$-$28386/EUCL$-$Q1$-$CL$\,$J033730.18$-$283827.6 (nicknamed the `Puddle' cluster), a $z\sim 1.74$ galaxy cluster candidate with a merging proto-BCG.

\begin{itemize}
    \item The Puddle cluster is a $z_\mathrm{phot}=1.65\pm 0.08$ candidate cluster in MaDCoWS2 catalogue and is detected as a $z_\mathrm{phot}\gtrsim 1.5$ candidate cluster by both \textit{Euclid} cluster-finding algorithms. Compared to field level, there is an excess of $108\pm 14$ galaxies brighter than $\HE=22.25$ within a $\ang{;2}$ radius around the proto-BCG. With a completeness correction, this number rises to $110\pm 14$ galaxies. This excess of galaxies strongly suggests that this candidate cluster is a genuine high-redshift overdensity, even though only the BCG is spectroscopically confirmed at $z_\mathrm{spec}\sim 1.74$ with H$\alpha$ emission.
    \item The Puddle cluster central \ang{;2} does not appear to contain luminous ($\HE\lesssim 21$) red galaxies aside from the BCG complex. We estimate the fraction of red galaxies ($\HE<22.25$ and $z-\HE>2.1$) as $17\pm 4$\% without the completeness correction and as $18\pm 4$\% with it. It should be noted however that $\HE<22.25$ does not correspond to a uniform stellar mass -- deeper $z$-band data would be needed to compute stellar-mass based red fractions.
    \item The proto-BCG appears to be a complex merger at $z_\mathrm{spec}\sim 1.74$, including 6 to 7 galaxies. Along its longest axis, the BCG system measures about $105\,$kpc, including its southwestern tail of diffuse emission. The emission spectrum associated with this system suggests that the brightest of the merging cores hosts an AGN.
    \item The results of SED modelling of the forming BCG indicate that the dominant stellar population is young ($0.32\pm 0.03\,$Gyr) and consistent with a short burst of star formation. The central AGN contribution to the SED is negligible. However, the most likely SEDs show substantial dust attenuation. Thus, more data are needed to determine the amount, if any, of obscured star-formation activity. The proto-BCG has a stellar mass of $(5.7\pm 0.3)\times 10^{11}\,M_\odot$.
    \item The morphology, size, and stellar mass of the proto-BCG, as well as the timescale of the star formation burst, suggest that the Puddle cluster could be a SPT2349$-$56-like merger, caught at a more advanced stage.
\end{itemize}

\begin{acknowledgements}
We thank C.C.~Steidel for providing an archival sensitivity function for MOSFIRE. A.T., A.H.G., S.A.S., S.T., D.S., P.R.M.E and B.M. acknowledge the support of NASA ROSES Grant 12-EUCLID11-0004. The work of D.S. and P.R.M.E. was carried out at the Jet Propulsion Laboratory, California Institute of Technology, under a contract with the National Aeronautics and Space Administration (80NM0018D0004). B.M. acknowledges funding support from the Jet Propulsion Lab (JPL) Euclid project.
\AckQone
\AckEC
This project used data obtained with the Dark Energy Camera (DECam), which was constructed by the Dark Energy Survey (DES) collaboration. Funding for the DES Projects has been provided by the U.S. Department of Energy, the U.S. National Science Foundation, the Ministry of Science and Education of Spain, the Science and Technology Facilities Council of the United Kingdom, the Higher Education Funding Council for England, the National Center for Supercomputing Applications at the University of Illinois at Urbana-Champaign, the Kavli Institute of Cosmological Physics at the University of Chicago, Center for Cosmology and Astro-Particle Physics at the Ohio State University, the Mitchell Institute for Fundamental Physics and Astronomy at Texas A$\&$M University, Financiadora de Estudos e Projetos, Fundação Carlos Chagas Filho de Amparo, Financiadora de Estudos e Projetos, Fundação Carlos Chagas Filho de Amparo à Pesquisa do Estado do Rio de Janeiro, Conselho Nacional de Desenvolvimento Científico e Tecnológico and the Ministério da Ciência, Tecnologia e Inovação, the Deutsche Forschungsgemeinschaft and the Collaborating Institutions in the Dark Energy Survey. The Collaborating Institutions are Argonne National Laboratory, the University of California at Santa Cruz, the University of Cambridge, Centro de Investigaciones Enérgeticas, Medioambientales y Tecnológicas–Madrid, the University of Chicago, University College London, the DES-Brazil Consortium, the University of Edinburgh, the Eidgenössische Technische Hochschule (ETH) Zürich, Fermi National Accelerator Laboratory, the University of Illinois at Urbana-Champaign, the Institut de Ciències de l'Espai (IEEC/CSIC), the Institut de Física d'Altes Energies, Lawrence Berkeley National Laboratory, the Ludwig-Maximilians Universität München and the associated Excellence Cluster Universe, the University of Michigan, the National Optical Astronomy Observatory, the University of Nottingham, the Ohio State University, the University of Pennsylvania, the University of Portsmouth, SLAC National Accelerator Laboratory, Stanford University, the University of Sussex, and Texas A$\&$M University. This work is based in part on observations made with the \textit{Spitzer} Space Telescope, which was operated by the Jet Propulsion Laboratory, California Institute of Technology under a contract with NASA. This research has made use of the NASA/IPAC Infrared Science Archive, which is funded by the National Aeronautics and Space Administration and operated by the California Institute of Technology. Some of the data presented herein were obtained at Keck Observatory, which is a private 501(c)3 non-profit organization operated as a scientific partnership among the California Institute of Technology, the University of California, and the National Aeronautics and Space Administration. The Observatory was made possible by the generous financial support of the W. M. Keck Foundation. This research has made use of the SVO Filter Profile Service `Carlos Rodrigo', funded by MCIN/AEI/10.13039/501100011033/ through grant PID2023-146210NB-I00.
\end{acknowledgements}

\bibliography{ref}

\label{LastPage}
\end{document}